# Alexa doesn't have that many feelings: Children's understanding of AI through interactions with smart speakers in their homes

**Valentina Andries**
Centre for Research in
Digital Education
The University of Edinburgh
valentina.andries@ed.ac.uk

**Judy Robertson**
Centre for Research in
Digital Education
The University of Edinburgh
judy.robertson@ed.ac.uk

As voice-based Conversational Assistants (CAs), including Alexa, Siri, Google Home, have become commonly embedded in households, many children now routinely interact with Artificial Intelligence (AI) systems. It is important to research children's experiences with consumer devices which use AI techniques because these shape their understanding of AI and its capabilities. We conducted a mixed-methods study (questionnaires and interviews) with primary-school children aged 6-11 in Scotland to establish children's understanding of how voice-based CAs work, how they perceive their cognitive abilities, agency and other human-like qualities, their awareness and trust of privacy aspects when using CAs and what they perceive as appropriate verbal interactions with CAs. Most children overestimated the CAs' intelligence and were uncertain about the systems' feelings or agency. They also lacked accurate understanding of data privacy and security aspects, and believed it was wrong to be rude to conversational assistants. Exploring children's current understanding of AI-supported technology has educational implications; such findings will enable educators to develop appropriate materials to address the pressing need for AI literacy.

**Key Words and Phrases:** Education; Conversational Assistants; Smart Speakers; AI education; Child-Computer Interaction; Trust; Anthropomorphism

## 1. INTRODUCTION

Artificial Intelligence (AI) and Large Language Models (LLMs) such as Chat GPT-3 and its successors have rapidly advanced in recent months, transforming our daily lives in unprecedented ways. At the time of writing (April 2023), there is considerable media attention focused on the advances of LLMs. Open AI has claimed that GPT4 can "reasonably be viewed as an early (yet still incomplete) version of an artificial general intelligence (AGI) system" (Bubeck et al, 2023). Citing concerns about an "out-of-control race to develop and deploy ever more powerful digital minds that no one – not even their creators – can understand, predict, or reliably control"[1], several thousand AI researchers have published an open letter calling for "all AI labs to immediately pause for at least 6 months the training of AI systems more powerful than GPT-4". Some commentators have criticised claims made by AI companies on the grounds that they are potentially inflated, lack transparency and disregard due scientific process (Marcus, 2023).

In a time of rapid change, against a backdrop of remarkable media claims and warnings about the capabilities of LLMs, members of the public could understandably be confused about what AI can and cannot do. While children may not themselves follow the media stories on this topic, the adults around them – parents, carers and teachers – may discuss them. But what do children already know about AI and how can they make sense of the advances which are likely to have a profound impact on their futures? In this research, we investigated children's understanding and opinions of a form of AI which is already firmly embedded in their lives.

---

[1] https://futureoflife.org/open-letter/pause-giant-ai-experiments/ /

This paper examines one popular form of AI-supported technology - voice-based Conversational Assistants (CAs) - which are commonly used by both adults and children in the home. These systems (such as Amazon's Alexa or Apple's Siri) offer a wide range of benefits which include educational support, and entertainment, while having positive implications for accessibility, especially with regards to touch-free interactions. Although not originally developed as 'social companions', CAs may be perceived as such because of their features which are aimed at providing a better user experience while building trust and emulating human-human interactions (Luger & Sellen, 2016). As many children use voice-based CAs regularly given their exposure to smart speaker devices such as Amazon's Alexa in the home, their perceptions and experiences give a window into the extent to which they perceive CA interactions to be human-like and trustworthy. It is possible that the features which are designed to build trust and acquire credibility via agent persona design (Braun et al., 2019; Volkel et al., 2020) are counterproductive for child users because they may be misled by the anthropomorphic affordances of the interface (Van Brummelen et al., 2021), causing them to overestimate the reliability and capabilities of the system with which they are interacting. This 'fundamental over attribution error' (Marcus and Davis, 2019) is likely to become a more pressing issue with rapid advances in generative language models such as ChatGPT. The problems arising from people's difficulties in distinguishing fact from fiction in ChatGPT's highly plausible responses are already being discussed in various media reports[2].

According to the Ofcom (2022) annual survey report on children's media use in the United Kingdom, conversational assistants in consumer devices (referred to as 'smart speakers' in the report) have been increasingly embedded into UK households in recent years. Smart speakers are used by children (aged 3 - 17) primarily for listening to music and acquiring information. The biggest increase in the UK was reported in 2019 by Ofcom, when the use of CAs by children aged five to fifteen almost doubled from 2018 to 2019. Compared to the Ofcom report released in 2021, covering the 2020-2021 period of time, children's access to CAs has increased from six in ten children to more than eight in ten children in 2022 (Ofcom, 2021; Ofcom, 2022).

Because children are spending more time interacting with AI, researchers have started investigating their experiences and implications of interactions with CAs (Druga et al., 2017; Druga et al., 2018; Druga et al., 2021; Van Brummelen et al., 2021). Most research on CAs and children remains US-centric, and only a few studies have been conducted in Europe to date (Druga et al., 2019; Festerling & Siraj, 2020; Fitton et al., 2018; Porayska-Pomsta et al., 2018). A systematic review conducted by Garg et al. (2022) highlights themes and trends that have been researched in the past decade in relation to CAs and children. Studies have focused on children's use of CAs in the home environment to elicit their attitudes towards these agents, as well as the children's understanding of this AI-supported technology. Such research has revealed children's perceptions of conversational assistants' traits, such as intelligence or anthropomorphic qualities.

In this study, we aimed to further investigate children's knowledge and perceptions of conversational assistants, by exploring Scottish children's understanding of these AI-supported systems. This was the first stage in co-developing a range of educational materials that can be used by primary-school teachers within Scotland to educate children on how CA technology works, the societal implications of its use, as well as foster AI literacy. Given the current climate of AI advances, it is important to investigate children's perceptions of the intelligence and anthropomorphic qualities of AI-supported conversational agents further, as well as in different countries and contexts. Moreover, as the literature is sparse with regards to young children's understanding of CAs' privacy aspects or their perceptions of what may be appropriate verbal interactions with CAs, an additional goal of the research reported here was to establish the extent to which children can accurately assess privacy risks associated with conversational assistants embedded in smart speakers in the home. These aspects constitute salient avenues for research, with direct implications for children's lives, and timely educational value to AI literacy (Long & Magerko, 2020; Ng et al., 2021; Ottenbreit-Leftwich et al., 2021).

---

[2] See for example https://www.ft.com/content/e34c24f6-1159-4b88-8d92-a4bda685a73c

## 2. RELATED WORK

Conversational agents (CAs) are systems that can participate in dialogues with users using either text or speech-based inputs and outputs. For voice-based CAs, the crucial elements include automatic speech recognition (ASR) and natural language understanding (NLU) components to comprehend user intentions, dialogue management to keep track of the dialogue history and state, and the ability to select an appropriate dialogue action based on these components. The language output is generated by a language generation component (NLG) and then converted to speech using text-to-speech synthesis (TTS) components. In comparison, text-based chatbots follow a similar design but rely solely on text inputs and outputs (Garg et al., 2022).

In the academic literature and in the media, this technology has been named in various ways, such as *conversational agents*, *conversational AI*, *digital voice assistants* or *virtual assistants*, but their common purpose is to assist the users by completing tasks assigned by them in real-time while also acquiring enough information about the users to mimic agency exertion on their behalf (Heuwinkel, 2013). In this paper, we focus on commercially available voice-based CAs which rely on AI (such as Alexa developed by Amazon). In our study, we used the term "smart speakers" to refer to CAs as an easier alternative for the younger children who participated in our studies, as well as to maintain a certain level of consistency with the OfCom (2021, 2022) reports and the terminology employed in their questionnaires addressed to children and their parents in the UK.

### *2.1. Children's understanding of CAs – cognitive abilities, agency, and anthropomorphism*

Anthropomorphism refers to humans' interactions with non-human entities, how they feel about such entities, or how they perceive them (Festerling & Siraj, 2022). CAs have the ability to communicate with humans in anthropomorphised ways, such as responding to queries, telling jokes and even thanking children for being polite in their interactions, which is a cue for children's perception of agency in non-living beings (Flanagan et al., 2023). Children may show a preference towards interacting with personified systems (Yuan et al., 2019), which have been given names, and they are more likely to regard systems which interact via natural speech as social agents (Strathmann et al., 2020; Yarosh et al., 2018).

According to the Human-Robot Interaction (HRI) literature, children, particularly younger children, have a natural tendency to assign human-like attributes to robotic or AI-powered entities. These attributes refer to children believing that robots display cognitive, moral, social or emotional states (Girouard-Hallam et al., 2021; Reinecke et al., 2021; Sommer et al., 2019). This relates to children's moral treatment of robots, such as children believing that is wrong to cause harm to robotic dogs (Melson et al., 2009), or to engage in harmful behaviours towards CAs or technologies such as the Roomba vacuum cleaner (Flanagan et al., 2023; Reinecke et al., 2021). According to Kahn et al. (2012), users perceive voice-supported systems as morally accountable, as opposed to simpler technological entities which do not have voice interfaces, such as a vending machine.

Anthropomorphising conversational assistants can present both dangers (Abercrombie et al., 2021) and benefits, particularly when it comes to children (Xu et al., 2021). A recent meta-analysis by Festerling and Siraj (2022) provides an overview of the positive implications of infusing technology with human-like qualities including encouraging children to explore and learn more through enjoyment and engagement (Papadopoulos et al., 2020; Xu et al., 2021), enhancing the effectiveness of collaborations between humans and technology, encouraging socially desirable behaviour from users (Bartneck & Hu, 2008; Festerling & Siraj, 2022; Shah et al., 2011; Złotowski et al., 2014) and eliciting a more durable sense of trust from humans (de Visser et al., 2016).

Anthropomorphising technology should however be tackled with a degree of caution, given the potentially adverse effect of imbuing systems with human-like attributes, also known as the *uncanny valley* (Mori, 2012). Users may react negatively to the uncanny likeness of an artificially intelligent system to humans, while children may perceive this to be 'creepy' (Yip et al., 2019). There are also ethical implications because anthropomorphisation could be regarded as a form of deception, potentially leading to children over-estimating the capacities of the AI-supported systems that they interact with (Sharkey & Sharkey, 2010). There is also a potential for developing social bonds with conversational assistants (Turkle, 2018), or other such types of interactions which may be regarded as reserved for human-human engagements (Sætra, 2020).

The perceived life-like qualities and intelligence of conversational agents appear to influence children's awareness of these systems' agency and animacy (Turkle, 2005; Xu and Warschauer, 2020). Previous research has also shown that by engaging in playful interactions with CAs, young children may regard agents as being friendly and truthful (Druga et al., 2017). Young children might overestimate the intelligence levels of such technologies by assuming that CAs are smarter than them because they 'know' plenty of facts (Lovato et al., 2019) or information which they deliver promptly to the user (Festerling & Siraj, 2020) and regard them as reliable sources of information by age seven (Girouard-Hallam & Danovitch, 2022). Children also assign anthropomorphic qualities to them because of agent personification or their speech style (Druga & Ko, 2021; Festerling & Siraj, 2020) thus misunderstanding the capabilities of such technologies and how they work.

While children attribute several human-like characteristics to conversational assistants and regard them as having mental, moral and social qualities such as intelligence and the ability to form friendships (Girouard-Hallam et al., 2021), children also acknowledge that these devices are not alive and cannot breathe (Girouard-Hallam & Danovitch, 2022). Thus, children do not view conversational assistants as either solely human or as inanimate objects but rather as combining characteristics from both categories (human-like and non-human) simultaneously, similarly to children's perceptions of humanoid robots' attributes (Festerling & Siraj, 2020; Girouard-Hallam & Danovitch, 2023; Kahn et al., 2012). This may suggest that a new ontological category has been emerging in human-robot interactions, and it extends to conversational assistants as well (Kahn et al., 2012; Girouard-Hallam & Danovitch, 2023). As more personified systems are created and increasingly adopted in daily life, this category may continue to evolve.

Children's perceptions of social and moral attributes change with age, with older children less inclined to assign social and emotional qualities to the CAs (Girouard-Hallam et al., 2021). This could be partly because younger children tend to develop para-social attachments with media characters from movies or games (Richards & Calvert, 2016), or because they become more aware of the limitations of the systems as they grow older. Although children may not perceive conversational assistants as fully autonomous entities, as long as children view these AI-supported systems as trustworthy (Druga et al., 2017) or consider them as possessing moral characteristics to some extent, they are prone to sharing excessive personal information with internet-connected devices (Girouard-Hallam et al., 2021). This could impact their internet safety and potentially mislead children to believe that these systems are reliable entities that can be trusted to keep user data and interactions secure.

This highlights a need to further research children's perceptions with the ultimate goal of developing their AI literacy. It has been established that it is necessary to teach children computer science starting at an early age (Robertson et al., 2017). Nowadays, as children's engagement with AI-supported technology is bound to increase, it is also becoming important to prepare these generations of children to understand how AI systems work, the potential of such technology that is integrated into everyday devices, as well as its limitations (Druga & Ko, 2021), and wider society implications (Ng et al., 2021). As Marcus and Davis comment: "It is increasingly important that we be able to sort out AI hype and AI reality, and to understand what AI currently can and cannot do" (Marcus & Davis; 2019; p24).

## *2.2. Trust, privacy and data management*

A systematic review published by Garg et al. (2022) provides an overview of themes of interest in researching children's use of CAs in the past decade in HCI literature. According to this, privacy issues remain important to investigate regarding such technologies that can seamlessly blend into the background in households and within easy reach to children. The review highlights that only very few studies have researched the concepts of children's data privacy and data management in relation to the information collected by such systems.

Children are regarded as particularly vulnerable users of online services and internet-connected products, and their online interactions with smart devices need to be protected and monitored. In some countries, specific laws have been passed to protect children online, such as the Children's Online Privacy Protection Act (COPPA) in the United States, the European Union's proposed Artificial Intelligence Act (EU, 2021) or some articles of the General Data Protection Regulation (GDPR) in Europe. There are privacy concerns surrounding children's use of conversational assistants, as the threats posed by these are similar to those raised by internet-connected toys which have raised data security concerns (Bicchierai, 2015; McReynolds et al., 2017; Valente & Cardenas, 2017). Zhang et al. (2017) showed that CAs such as Alexa, Siri or Cortana can be manipulated by hackers by injecting commands towards these systems, which are imperceptible to human users. Such commands can then be utilised by hackers to make phone calls, change phone settings, or to manipulate home security.

The companies which manufacture consumer devices may also employ questionable approaches to data management and privacy, potentially leading to breaches in security. In 2015, an attack on the servers of the Chinese VTech company which sells children's devices (phones, tablets, baby monitors) revealed that personal information from approximately five million parents and 200,000 children had been stored on those devices (Bicchierai, 2015).

Such data security malpractice may have implications for third-party applications available for daily use with Amazon's CA, Alexa; these are also known as *Skills*. With over 100,000 skills available for users to order items, turn on lights in their house, check their credit card balance, or access entertainment options such as trivia, impromptu translations and storytelling for children (Kozuch, 2023), there are concerns about the privacy and security of personal data required to enable such services (Sabir et al., 2022). Although Amazon has security procedures in place for vetting third-party services to ensure the safety of the applications that would be linked with Alexa, cyber-security research has identified various gaps in their vetting skills process. For instance, a study by Lentzsch et al. (2021) found that 23% of the 1146 Alexa skills that they investigated and which requested access from users to different types of personal data (which could be name, phone number, geo-location etc.) did not have privacy policies in place or they were either incomplete or misleading, such as not fully disclosing all types of sensitive data accessed (Liao et al., 2020).

Furthermore, Lentzsch et al. (2021) showed that it is possible for developers to modify the code on the backend of skills even after the skills have been made available for download. To illustrate, they created a skill and subsequently made changes to the code to request more user information, despite the skill having already been approved by Amazon. Lack of security vetting, in addition to the lack of security indicators which would help users identify which skills are provided by third parties and which use personal data, leads to ambiguity in user data practices (Sabir et al., 2022). This raises questions of trust in AI-supported systems, and leaves users, especially children, at risk of having personal data shared in ways to which they may not have consented.

Some of these privacy concerns are currently echoed in the United Kingdom by the most recent Ofcom (2022) report on the media use and attitudes of children and parents. The survey sampled 2500 parents of children aged three to seventeen-years old, who responded to Ofcom's questions related to attitudes, concerns and mediation strategies for children's media use. The highest proportion of parental concerns were related to the device misunderstanding the child's instructions and responding with inappropriate content (37%) or that the children would specifically enquire about inappropriate content themselves (31%). Their other concerns revolved around privacy and data management: 28% of the parents showed reluctance to the child's use of CAs due to the device collecting recordings of interactions, collecting children's data (23%), or more generally being worried about the device listening to the children's conversations (23%).

Data management (or mismanagement) and system security for child users of smart speakers need to be researched at length and discussed with children, parents and other relevant stakeholders. This would help increase their awareness of technological vulnerabilities, potentially having direct implications for their use of smart technology. The importance of this topic is echoed by a recent report produced by UNICEF (2022), which highlights the fact that children's rights extend to digital rights that include respecting children's privacy when surfing the internet or using internet-connected services.

## *2.3. Interactions with CAs*

The use of conversational assistants like Alexa raises questions about how to treat them given that they can sometimes be perceived both as social companions and inanimate artefacts (Burton and Gaskin, 2019). Since conversational assistants' performance is not dependent upon users' politeness, users make brief or abrupt requests without using polite language or even employing offensive terminology at times (Chin et al., 2020; Curry and Rieser, 2018). A study by Veletsianos et al. (2008) with teenagers aged 14-15 in the United States focused on interactions with a text-based pedagogical assistant. They found that over 40% of the pupils' interactions with the assistant were inappropriate, ranging from adopting impolite attitudes to offensive language, as well as sexually explicit terminology.

This has raised concerns about whether this behaviour towards AI-supported systems is making people and children less polite in their human-human interactions (Burton & Gaskin, 2019; Gartenberg, 2017; Hiniker et al., 2021). Some parents are concerned about the impact on their children's manners in daily interactions and Amazon has responded by adding a "magic word" feature that rewards children who ask politely (Baig, 2018; Truong, 2016).

According to research primarily with adult users, both text-based and voice-enabled CAs are susceptible to verbal abuse (Branham, 2005; Chin et al., 2020; Curry & Rieser, 2018;

Curry & Rieser, 2019), which could potentially normalise abusive behaviour in real life. For instance, previous research with adult users has shown that repeated exposure to simulated violence, such as abusing a robot, can desensitise individuals to violent actions in real life (Whitby, 2008).

In a study by Festerling and Siraj (2020) on young children's use of CAs in Germany (N=27; ages 6-10), insights about verbal interactions emerged as part of the authors' discussion with the children and observation of their interactions with CAs in focus groups. They noted that some children used friendly and affectionate language when interacting with the CAs, as well as showing polite manners by using terms such as "please" and "thank you". However, more children would use unfriendly or even aggressive terms (and dominant body language) to interact with the CAs (e.g. "you stupid thing" while hovering over the device).

This may suggest that children may perceive these agents as less worthy of respectful language in comparison to humans or living entities. From this perspective, children may perceive voice assistants such as Alexa and Google Home as morally inferior to humans (Festerling & Siraj, 2020), viewing humans as the creators of machines, while machines have inherent limitations that prevent them from surpassing human intelligence. According to this study (Festerling & Siraj, 2020), children view the CAs' permanent responsiveness as a non-humanoid characteristic and perceive AI systems as assistants created to always respond to users' demands and queries, while humans have the moral right to remain silent or disregard others if they wish.

However, the authors' inferences about children's views of the moral standing of AI systems contrast with the findings of Flanagan et al. (2023). They employed a questionnaire approach to investigate 127 (4-11-year-old) children's perceptions of the Alexa CA, the Roomba vacuum cleaner and the small humanoid-looking Nao robot. All the participants in their study believed that the systems have moral standing, and consequently, that it was wrong to harm any of the technologies that they were asked about, irrespective of their function, abilities, humanoid/non-humanoid/disembodied appearance and movement capabilities.

Such study findings and insights suggest the need for further research on the topic of children's verbal interactions with AI-supported assistants, raising additional questions about what is considered appropriate in terms of interacting with smart systems, whether human users should be expected to use polite language in their verbal interactions with CAs, and potentially what may constitute appropriate verbal response strategies to mitigate users' offensive language (Chin et al., 2020; Curry & Rieser, 2018). Learning about children's perceptions of politeness norms towards CAs would provide further insights into their broader understanding of AI systems as entities endowed with (or lacking in) agency or anthropomorphic qualities.

## *2.4. Research questions*

The research questions are as follows:

(1) General knowledge and understanding of CAs: What do primary school-aged children know about Conversational Agents in terms of their cognitive abilities, perceived intelligence, anthropomorphic features, and how this AI technology works?
(2) Data privacy concerns: What do primary school-aged children know about Conversational Agents and data privacy, and to what extent do they trust them?
(3) Interactions with CAs: How do children interact with CAs? What do young users consider to be appropriate verbal interactions with Alexa?

## 3. METHODOLOGY

We employed a mixed-methods approach in this study involving primary-school children in Scotland in 2022. We used questionnaires with 8-11-year-old primary-school children, and group interviews (6-11 years old). According to the review published by Garg et al. (2022) on CAs and children, only a small amount of the papers included in their review employed a mixed-methods approach (5 papers out of 38), the majority of the studies having adopted either a quantitative or a qualitative approach. We aimed at fostering broader participation in our study by means of the questionnaire approach, capturing a wider range of children's answers, whereas the interviews allowed for in-depth discussion of the main questions, the different methods providing complementary insights into the research questions.

The methodology was thoroughly discussed with specialists in education from the University, as well as with teaching staff from the primary schools from which participants were recruited. This ensured that the study approach was sensitive to the participants' needs, the context for data collection, and the gatekeepers who would facilitate recruitment, while also allowing for flexibility to adapt the approach in light of changing pandemic regulations.

After discussion with the teachers and pilot participants, we chose to use the terms 'smart speaker' and 'Alexa' to refer to conversational assistants deployed on home devices. This was to make the questionnaire questions more readable for children with lower literacy levels. The teachers advised that the term 'Alexa' is familiar to children. A recent OfCom report (2022 reported that Amazon's Alexa is the most commonly owned smart speaker in the UK; this was borne out by the very high proportion of our participants who own this device.

### *3.1. Recruitment*

We liaised with school staff (teachers and head teachers) from two primary schools in central Scotland for recruitment purposes. One school is located in a village and has a postcode corresponding to the 6$^{th}$ and 7$^{th}$ least deprived deciles of the Scottish Index of Multiple Deprivation (SIMD)[3]. The other school is in the suburbs of a city, with a postcode in the least deprived 10% according to SIMD. The teachers were the gatekeepers who provided access to their own classes as well as to children from other grades by liaising with their colleagues for the purposes of this study.

Ethical approval was obtained from the University, having prepared participant information sheets and consent forms for both the parents and their children. We liaised with primary-school teachers for ethics approval purposes as well, to distribute these forms to the parents and the children, either online (using Microsoft Forms) or by providing paper copies. As part of the ethical approval process, a COVID-19 risk assessment had to be completed by the first author, who also carried out the data collection.

### *3.2. Questionnaires*

The questions that we included in the questionnaire were selected after consulting the HCI, human-robot interaction (HRI) and developmental psychology literature on investigating children's understanding and perceptions of conversational assistants, smart toys, robotic technology and AI. See Appendix 1 for a list of the questionnaire and interview questions with references to the studies from which they were drawn including Robertson et al. (2017); Melson et al. (2009); Van Straten et al. (2020); Van Brummelen et al. (2021).

The questionnaire questions were piloted in an in-person informal consultation with children from grade P5 (N=7; age=9) together with their teacher from one of the schools that later took part in the formal study. The questionnaire was distributed to two different primary schools in the autumn of 2022. The questionnaire was completed during class time in either digital or paper format, depending on device availability. The completion time was between 5 to 10 minutes, depending on the children's age and possible assistance that was needed. The children did not require much help to read, understand or answer the questions; the assistance provided by the teachers was primarily related to spelling. The online version of the questionnaire was done using the Qualtrics platform; this approach was feasible in the case of the children who had access to iPads or computers in school. 166 children completed the questionnaire, with the demographics shown in Table 1.

---

[3] https://simd.scot/#/simd2020/BTTTFTT/9/-4.0000/55.9000/

| Gender/Age | 8 years | 9 years | 10 years | 11 years | Total |
|---|---|---|---|---|---|
| Male | 17 | 25 | 19 | 15 | 76 |
| Female | 30 | 16 | 14 | 17 | 77 |
| Prefer not to say | 8 | 4 | 0 | 1 | 13 |

*Table 1. Demographics of questionnaire participants*

### 3.3. Interviews

A group interview approach was preferred for gaining an in-depth yet interactive understanding of children's perceptions of CAs. A total of 10 group interviews (N=28) was conducted by the first author with children from primary-school years 2 (age 6), 3 (age 7), 4 (age 8), 5 (age 9), 6 (age 10) and 7 (ages 11-12). The demographics are shown in Table 2. The majority of the interviews were conducted with three children per session. The grouping of the pupils was done primarily by age and class range, with the help of the relevant teaching staff who had knowledge of which children from their classes had a positive rapport with each other, while also ensuring that the participants met the inclusion criteria (diverse backgrounds, genders and abilities).

| Gender and Participant Numbers per Group/Group Numbers | G1 | G2 | G3 | G4 | G5 | G6 | G7 | G8 | G9 | G10 |
|---|---|---|---|---|---|---|---|---|---|---|
| Female (F)/ Male (M)/ Prefer not to say (PN) | F1 F2 M | M1 M2 F | F1 F2 | M1 M2 F | F1 F2 M | F M1 M2 | M1 M2 M3 | PN1 PN2 M | F1 F2 | F M1 M2 |
| **Primary School Year** | All P4 | All P5 | Both P4 | P5 (M1); P6 (M2; F) | All P6 | All P6 | All P7 | All P5 | P3 (F1); P2 (F2) | All P7 |

*Table 2. Demographics of interview participants*

The interview schedule followed a semi-structured approach which was considered highly suitable given the wider age range of the participants and the need to adapt the language accordingly. The interview questions (Appendix 1) allowed for a more in-depth investigation of the questionnaire data. The interviews lasted between 13 and 30 minutes.

The researcher clarified any misconceptions about the topics covered at the end of the interviews, to avoid disrupting the group interview interactions. The educational materials that we are currently developing together with children will be distributed to the classes that took part in the questionnaire and the teachers who facilitated the recruitment process.

### *3.4. Data Analysis*

Descriptive statistics and graphs were generated from the quantitative data using R (package likert). All graphs can be viewed in Appendix 2; to save space only selected graphs appear in the main text. The data from the semi-structured interviews was analysed qualitatively, using the NVivo software (version 12) for coding themes and sub-themes. As an initial step to analysis, the first author transcribed the audio recordings of the interviews, familiarising themselves with the data.

Thematic analysis was considered suitable for analysis as it allows for observing patterns and themes across data (Braun & Clarke, 2006). The interview scripts were printed and paper copies were used for the first step of the analysis, using colour coding and annotations. These codes were then added to the scripts in the NVivo analysis tool.

The analysis process started at the stage of collecting the data, following the concept of interweaving (Denzin & Lincoln, 2011). The initial themes were: the anthropomorphism of CAs, agency of CAs, cognitive abilities of CAs, trust in CAs and interactions with CAs. New themes and sub-categories were extracted at a later stage of the data analysis (such as misconceptions about how Alexa works and children's understanding of Alexa's architecture). These emergent themes were primarily used for the development of educational materials about smart speakers and will not be addressed explicitly in this paper. The codes were shared and refined with the other researcher, who also applied them to the data for inter-rater reliability purposes. The themes were discussed, some of the codes were modified and refined, while also further themes were iteratively developed.

## 4. FINDINGS

The findings of the questionnaire and interviews are reported below by theme.

### *4.1. Overview of children's use of Conversational Assistants*

The majority of the children who responded to the questionnaire reported that they had a smart speaker at home (92%). Amazon Alexa was the most commonly owned system (79%), followed by Google Home (11%), Siri (7%), and other systems (3%). Across all age groups, 44% of the children reported using their smart speakers very often, 45% sometimes used it, and 11% did not use it often. As shown in Table 3, the older children reported using their smart speakers less frequently.

| Frequency of use | 8 years | 9 years | 10 years | 11 years |
|---|---|---|---|---|
| Very often (daily) | 34.85% | 30.30% | 18.18% | 16.67% |
| Sometimes | 35.29% | 27.94% | 20.59% | 16.18% |
| Not often | 17.65% | 0.00% | 35.29% | 47.06% |

*Table 3 Frequency of smart speaker usage by age (Questionnaires)*

In the questionnaire, the most frequently selected activity with smart speakers was playing music (40% of the selections), followed by asking questions (20%), jokes (12%), searching for information (11%), listening to stories (6%), and other activities (6%). During the interviews, the children mentioned similar purposes for using their smart speakers, and also mentioned help with homework and controlling home devices such as lights.

Most of the children we interviewed found this technology to perform well in most cases, meaning that the information or answers provided by CAs were provided quickly and they were usually correct. Some of the interviewees also noted that their smart speakers would remember their preferences and even their names, e.g. *"it knows what I like – without having to think about what music to listen to" (G6, F)*.

Other children noticed the contrary, describing instances when their smart speaker would not perform as anticipated or requested, such as playing different songs than the ones requested due to multiple songs having similar titles, or the CA misunderstanding words from the song titles leading to no results. Some of the children also observed that their smart speakers seemed to understand their parents better than them. Nonetheless, they regarded this type of technology as generally high performing because of its perceived smartness, and overall helpful to have available, especially *"when my hands are full and we're eating" (G7, M1)*, recognising the usefulness of touch-free technology.

*4.2. Anthropomorphism and agency*

When asked explicitly to categorise technology like Alexa, 82% of the participants said it was 'Artificial Intelligence'. Only 1% said it was human, 1% an animal, 15% an object and 1% described it as 'other'. However, the questionnaire results relating to anthropomorphism and agency show an overall pattern of confusion, with around one-third to one-half of the participants indicating that they were unsure about whether smart speakers have feelings or agency. This suggests that while the majority of the children do not believe that smart speakers are literally human, they are confused about what to believe about the human-like behaviours which they exhibit.

The highest proportion of the participants were unsure whether their smart speaker has feelings ('maybe' 37%), while 35% of the participants believed that it does not, and 28% believed that it does. Interestingly, 50% of the sample were unsure as to whether their smart speaker was capable of feeling offended, 23% thought that it could, and 27% thought that it could not. It is not clear why there is a discrepancy here. It is possible that of all the variety of possible emotional behaviours, smart speakers' responses most often give the impression of being offended, e.g. by not responding to commands.

When asked whether their smart speaker has friends, 42% of the sample were unsure, 36% said 'no' and 22% answered 'yes'. The majority of participants (55%) thought that it was not acceptable to throw away their smart speaker if it was broken, with 32% unsure and 13% thinking it was acceptable. Note that the interview findings suggest that reluctance to discard a broken device might be due to sustainability concerns rather than an emotional attachment. Over half of the questionnaire respondents did not believe that smart speakers chose to do things (55%), with 32% unsure, and 13% of the sample believed that smart speakers do choose to do things.

The questionnaire results relating to anthropomorphism (broken down by age group) are shown in Figure 1. Overall, the largest proportion of the nine and ten year old children were unsure as to whether smart speakers have feelings (49% and 44% respectively), the largest proportion of the youngest and oldest children believed that they do not (40%). The younger children (aged 8 and 9) were more likely to be unsure as to whether smart speakers choose to do things than their older peers. The younger children in the sample were generally more inclined to think a smart speaker is capable of being offended, to believe that it might feel left out of conversations, and to consider it unacceptable to discard a broken smart speaker.

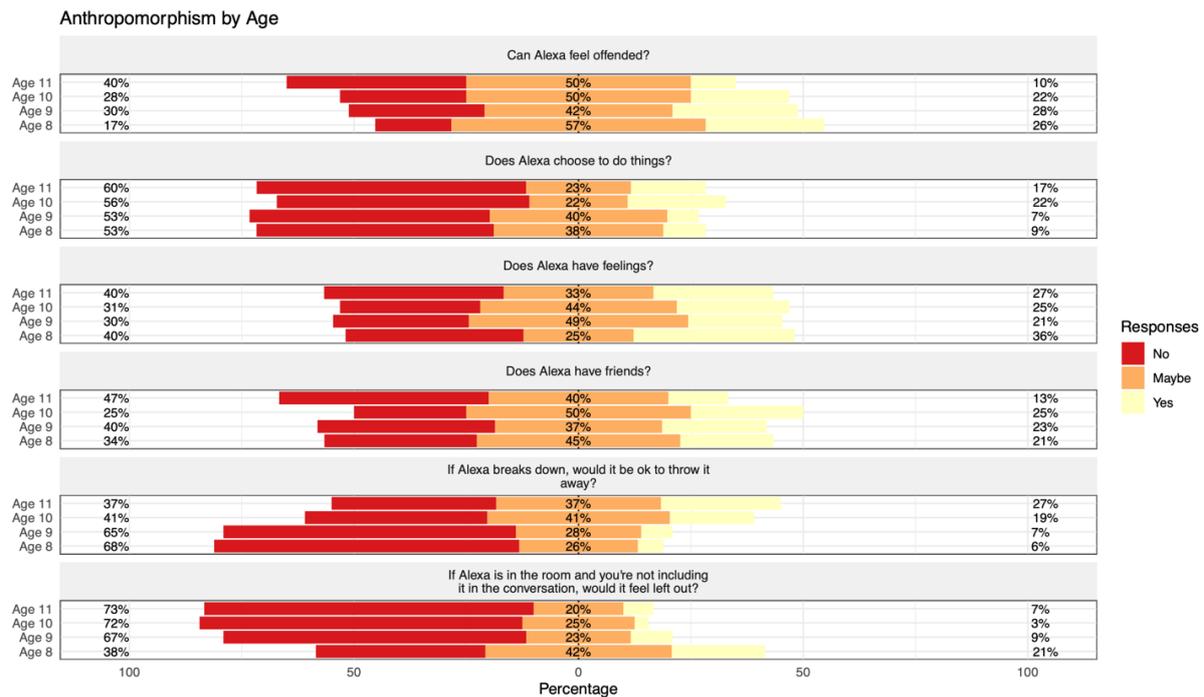

*Figure 1. Questionnaire questions relating to anthropomorphism by age*

From the interviews, many of the children expressed some doubt about whether smart speakers have feelings. The younger children were more inclined to attribute emotions, often based on the behaviour of the smart speaker. For example, some children thought that when Alexa sings happy birthday to the user that it is because it is happy, or that it goes silent when the user shouts at it because it is sulking. One girl expanded on this idea:

> "Alexa is your friend. She plays you music that you ask. She doesn't go, 'No, I don't want to', she just plays it. She if you're just mean to her, then she was alive, she'd be mean to you just like payback." (G5, F1)

Some of the children who were not sure whether smart speakers have feelings reasoned that it was better to behave as if it had feelings in case there was "someone inside" who could feel offended. One girl explained that she watched a video about animals' capacities to experience emotions which left her with the impression that humans underestimate animals' feelings; she, therefore, thought it was prudent to give smart speakers the benefit of the doubt. One child thought that smart speakers could have feelings "deep down" and that this could be a result of programming. The older children were inclined to say that smart speakers do not have feelings, although some speculated whether they could in the future:

> *"But as soon as there is … proper artificial intelligence that actually does have feelings - then I would say no, it shouldn't [continue to help when a user is rude to it]. It shouldn't because it has feelings to be hurt. But Alexa doesn't"* (G10, M1).

When asked whether they would replace their smart speaker with another one if it broke, some children did not want to because they felt an attachment to a particular device: "I have a Google and he's very special to me" (G10, M2); "If it broke we could have a gravestone for it in the garden" (G5, F1). For other children, their smart speaker simply provided a service and would be interchangeable with an identical device. It was common for the children to point out that it was not environmentally sustainable to throw away a device which doesn't work and that they would try to fix a broken smart speaker.

In general, the children did not attribute agency to smart speakers in the sense that they would do things of their own volition. This is summed up by this excerpt: *"Because it's not their own person. They don't have a brain for themselves or a heart or any feelings. It's just an AI made by someone else"* (G7, M2).

The interviewed children generally thought that the smart speakers followed instructions given to them by humans, or were directly controlled by humans. One group described it as humans and artificial intelligence working together. In some interviewees' opinions, smart speakers are designed as assistants to humans to help them carry out tasks or find information. One child speculated that smart speakers might be controlled by the company which makes them. Although the children did not generally attribute agency to smart speakers, some did believe them to be worthy of moral consideration:

> *"I'd say it's not human like us, but I'd say that you can't really treat it unfairly, like you can't chuck it down the stairs cause you also can't chuck a child down the stairs, right?"* (G6, F)

### *4.3. Cognitive Abilities*

There was a general pattern within the questionnaire items relating to cognitive abilities that the participants tended to overestimate the intelligence of their smart speakers, or at least were unsure about them. When asked whether Alexa always gives you the correct information, most of the participants were appropriately sceptical (21% 'no', 44% 'maybe'). 35% of the overall participants said 'yes' and this increased to 40% in the youngest age group (see Figure 2). Most children agreed that Alexa was smart (79% 'yes', 12% 'maybe', '9% no).

The highest proportion of the questionnaire respondents showed that they were unsure whether their smart speaker could think for itself (39% 'maybe'), with 32% indicating that it could not and 29% indicating that it could. The participants who selected the option that the smart speaker could think for itself were asked 'how much' it could do this – 37% thought that it could think for itself 'a little bit', 46% said 'a medium amount' and 17% said 'a lot'. Again, the highest proportion of the participants were unsure as to whether a smart speaker could make decisions (40% 'maybe', 27% 'no', 33% 'yes').

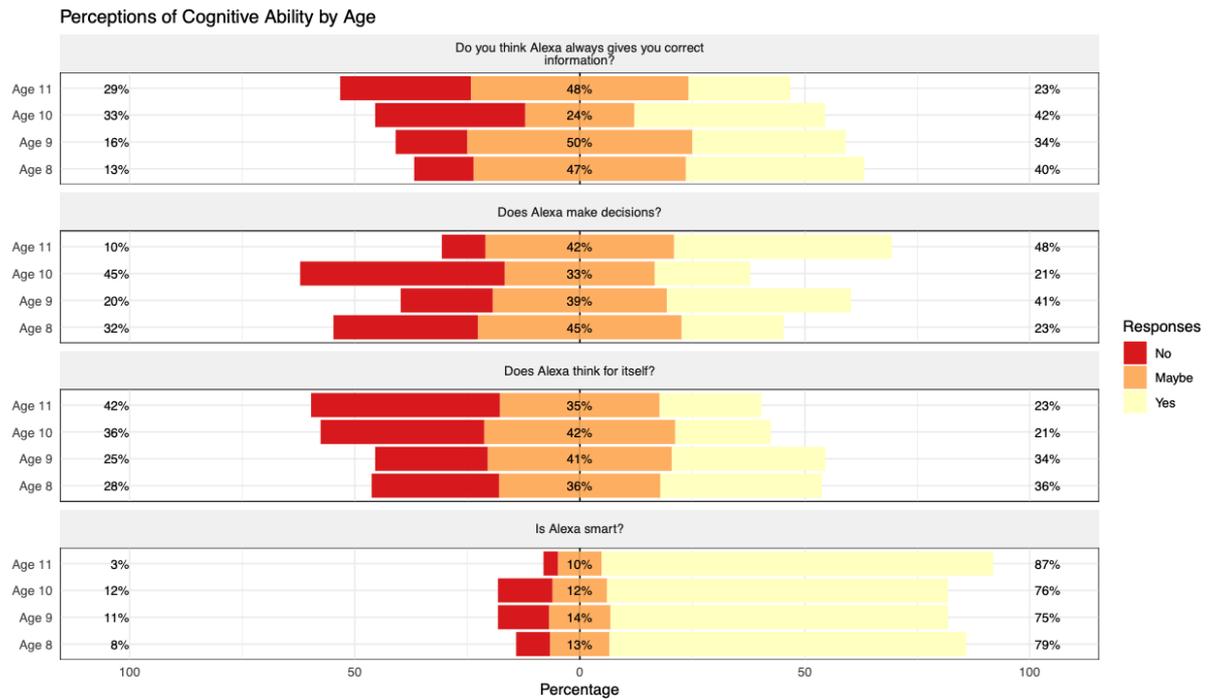

*Figure 2. Perceptions of smart speakers' cognitive abilities by age*

As shown in Figure 2, the younger children were more inclined to believe that smart speakers think for themselves (although the pattern is mixed for the question about whether they can think for themselves). The oldest children (11-year-olds) were more likely to say that Alexa is smart, even though they realised that the information from the device was not always correct.

For the interviewees, smart speakers generally featured on a notional scale of "smartness" somewhere above children but below scientists. Sometimes smart speakers were considered to be less clever than particular individuals such as older siblings or indeed themselves - *"[Alexa] is definitely smarter than you. Maybe not me" (G1, M)*. Reasons for thinking that the devices were smart included their ability to do difficult calculations quickly, answer questions on a wide range of topics, play music and create lyrics for rap songs. Interviewees often commented that smart speakers had access to human knowledge, for example, by searching the internet. "Being smart" was often (but not only) associated with having quick access to large quantities of facts, and some children identified that they, as children, were still in the process of learning this kind of knowledge.

> *"M2: It really depends what you're doing because like I asked, Alexa, how many people play Fortnite in the world? And Alexa answered it in like one second and scientists would take 10 minutes.*
>
> *F: Or maybe two years to actually pick up a survey" (G2)*

Some of the older children noted that the apparent cognitive abilities of smart speakers were programmed by scientists or software developers e.g. *"I don't think that Alexa is smart but the Alexa designers and coders are smart. Clever – and they used their smartness to cleverly program Alexa." (G10, M1)*. However, some of the children were puzzled as to how this might be achieved, as illustrated in this exchange:

> *"F2: Has it been like programmed - this will probably take a long time - for every question you could ask like, look at the answer and put it in the Alexa?*

> *F1: No, cause if you ask it about the weather then they would have needed to know the future to do that.." (G5).*

Interestingly, although F1 (G5) thought that Alexa was programmed to be smart, she was still puzzled by its cognitive abilities for its age - *"how does she know that much if she's 6 [years old]?"*. This suggests that some children could assume that machines undergo a developmental learning process in a similar way to humans.

When asked how smart speakers work, some children mentioned hardware components such as plugs, wires, hard drives, microphones and sensors but did not explain how these might be connected. Some children attempted explanations of how smart speakers might answer queries e.g. the following quote illustrates one of the more accurate explanations:

> *"[there are] microphones - and it detects you telling her something and when you say the word 'Alexa', she stops listening in a lot and then she's got a giant dictionary of audio examples from enough people so that however - whatever your accent is, in a particular language - then she'll be able to understand you. OK and then she's like, 'OK, the person said, 'Alexa, turn on kitchen lights', find in my personal Alexa database what kitchen lights are, and then it turns them on." (G10, M1)*

Other children believed that humans were more directly involved in answering queries:

> *"I think Alexa's made by humans who are on a computer and then they listen to what you're saying. And then when it says, 'Alexa', it turns on and then they type something in. So that's why Alexa doesn't think for itself" (G8, PN1).*

Another child thought that this task could be achieved by a single individual: *"Because of a really smart guy. He plugged himself into it and then it [Alexa] got really smart and then he plunged himself into multiple [Alexas] and now there's an army" (G1, M)*. As an extension of this idea, another member of the same group thought that a single human had shared his knowledge in advance by pre-recording all possible responses rather than answering live. *"I think someone spoke into a speaker and said every possible thing in the world and then he put in the devices" (G1, F1).* However, a different group of children identified that it would take "a while" to program all possible responses.

The children mostly responded with functional suggestions of what smart speakers do when they are switched off e.g. charging, updating, or learning more information. They speculated as to whether the device would be listening when it is switched off. However, none of the interviewees gave an answer to suggest that the smart speaker would be harmed if it was switched off. Some children used the analogy of the device going to sleep; one compared it to unconsciousness because it is harder to wake from.

Questionnaire participants were asked this as an open-ended question and the majority of questionnaire responses echoed those provided by the interviewees, e.g. recharging or sleeping ("It probably goes to sleep because it can't be bothered with us humans"), and some responses described smart speakers as doing nothing. Other responses indicated that the children were not sure or did not know what happens when smart speakers are off, and a small number or respondents said that it is "dead" but there was no elaboration on the topic. Similarly to the interview responses, some questionnaire respondents speculated that the speakers might (still) be listening despite being switched off:

> *"She [Alexa] lisens to my seacreat conversations"*
>
> *"It [Alexa] listen's to what your saying and tells the company that it works for "* (questionnaire respondents)

### *4.4. Data privacy and trust in Conversational Assistants*

Questionnaire participants were asked a series of questions relating to issues of trust. The majority of the participants said that they trusted their smart speaker less than they would a friend (62%), with 28% saying that they would trust it the same as a friend, and 9% more so. As shown in Figure 3, the youngest children were more likely to trust it the same (49%) as a friend or more than a friend (13%).

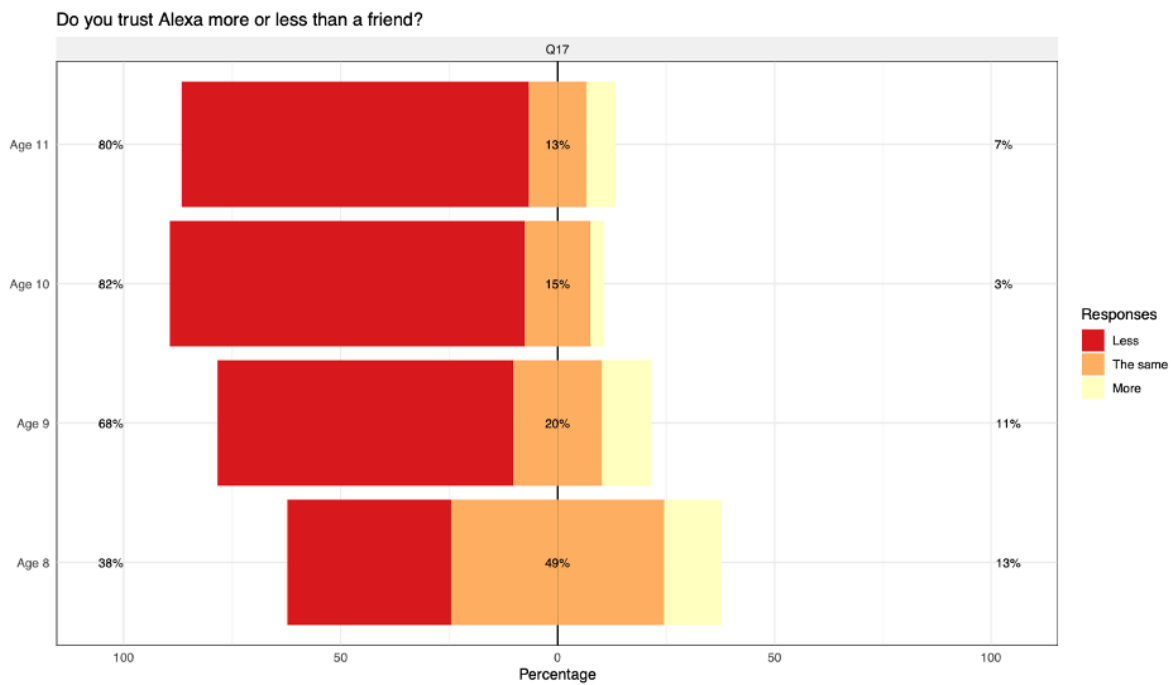

*Figure 3. Trust in smart speakers compared to a friend, by age*

Most of the children in this sample did not tell secrets to their smart speaker (82%) with only 9% agreeing with this or saying that they do so sometimes. Again, the younger children (aged 8 and 9) were more likely to exhibit trust by telling the smart speaker secrets.

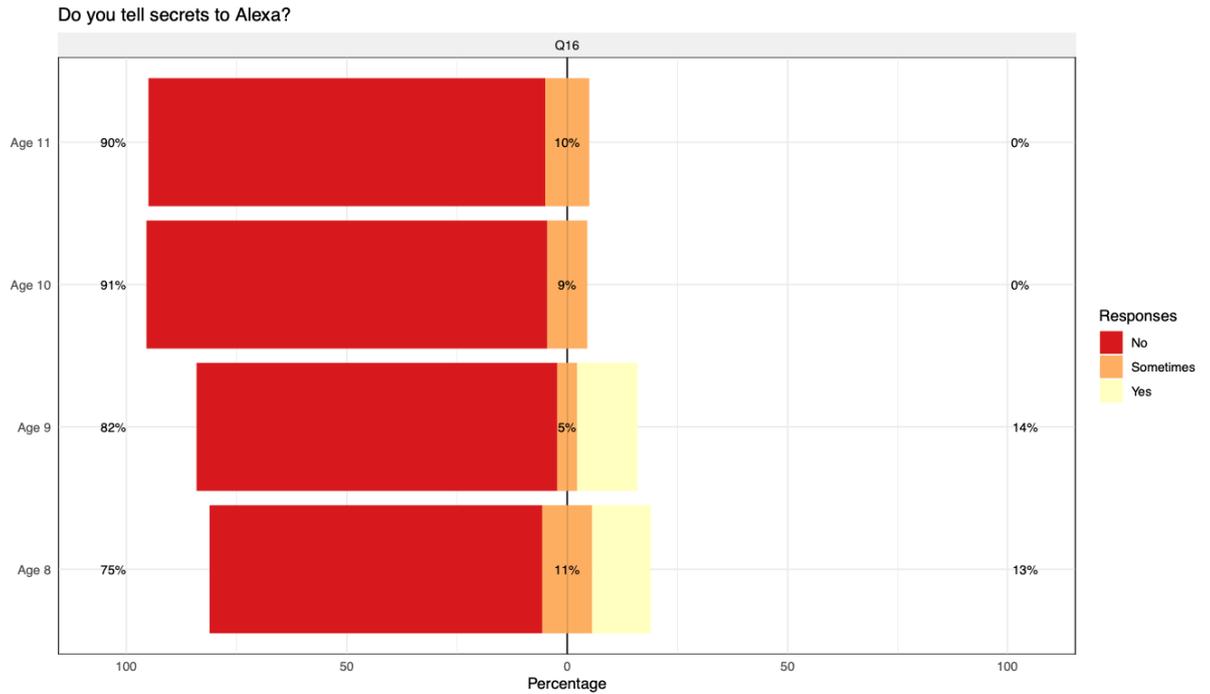

*Figure 4. Whether participants tell secrets to their smart speaker, by age*

Most of the participants reported that they would feel upset if strangers could listen to everything they said to their smart speaker ('yes' 71%, 'maybe' 12%, 'no' 17%). Fewer participants indicated that they would be upset if other people from their own home could listen to what they said to their smart speaker ('yes' 34%, 'maybe' 22%, 'no' 44%). The majority (61%) would feel comfortable if their smart speaker could recognise them by their voice ('maybe' 18%, 'no' 21%).

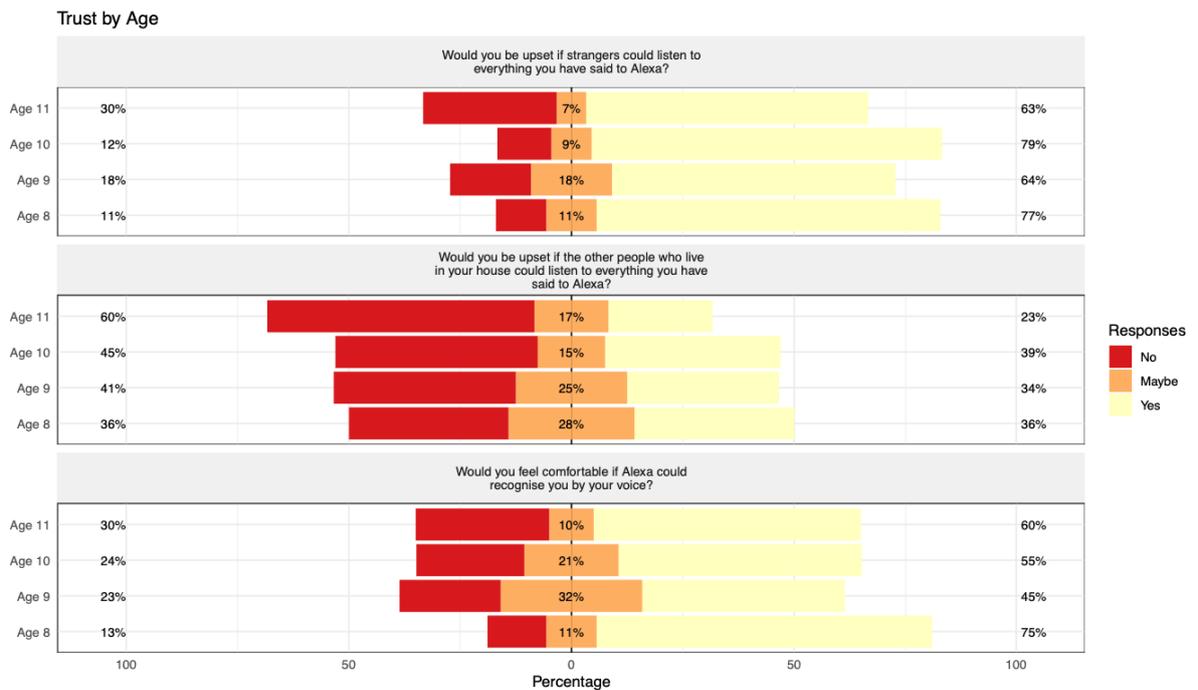

*Figure 5. Trust in smart speakers by age*

Figure 5 shows the patterns of attitudes to trust issues broken down by age group. The oldest children (11 years old) were less likely to be upset by interactions with the smart speaker being accessible to strangers or people at home. The youngest children (8 years old) were more likely to feel comfortable if the smart speaker could recognise them by voice.

Discussing with the children about their interactions with CAs in the interviews revealed two areas of interest: children's perceptions of data privacy and aspects related to the security of users' data.

Consistently to the questionnaire findings, when the children were asked whether they share secrets with their smart speakers, the majority said that they do not. Nonetheless, some of the children revealed that they told secrets to the smart speakers; this was usually motivated by their perceptions of CAs as trustworthy. Two children discussed that they trust their friends less than their smart speakers because their friends can tell secrets to others. Some children perceived their CA as a dehumanised system; this perception seemed to be beneficial for their personal interactions with CAs:

> *"Friends have mouths – Alexa does not" (G1, F2)*

> *"If one day I have like a really bad day so I start talking to Alexa because I just, honestly, I feel SO bad, I just need to talk to someone and I don't want to talk to a human" (G6, M1).*

The children who mentioned that they did not disclose secrets motivated their choice primarily by highlighting their lack of trust in such systems as they could somehow "tell" or "spread" their secrets. They were inclined to express a cautious outlook on more private interactions with CAs because of the perceived consequences of smart speakers mishandling their private information:

> *"M2: I feel like if I'm ever gonna tell secrets to Alexa, they could tell some other people, then they could tell some other people and ..*
>
> *M1: Then tell the whole world! If you tell Alexa secrets, I feel those messages travel to Alexas throughout the Alexa Amazon network and then if someone's telling their secrets to Alexa and someone else asks them to tell their secrets but then it tells YOUR secrets! Then I would not feel safe cause the whole world might know…It could be my friend that Alexa's travelled to but who knows…" (G6)*

Another group of children took this view slightly further, describing a potential conspiracy scenario involving CAs developed by different companies which may collaborate to spread private information:

> *"M1: So Alexa and Siri work together and then they could spread on one another […] and then could go viral*
>
> *M2: And then everyone will be able to know your personal secrets*
>
> *M1: And then you'll just you'll feel very sad*
>
> *F: And you'll be very embarrassed" (G2)*

One group of older children regarded the idea of sharing secrets with smart speakers as an impractical approach. Their perception of CAs as dehumanised entities was in this case detrimental to their personal interactions with CAs:

> *"M1: It's not like talking to a person where they will take it seriously. She'll just say 'Okay'.*
>
> *M3: 'I'm sorry, I don't know that one.'*
>
> *M2: … I really don't get the point. She doesn't even have feelings*
>
> *M1: Like your parents will probably do something about it, but Alexa will just give you adverts for like, anti-crying material or something" (G7).*

Most of the interviewees expressed negative reactions to the possibility of their interactions with CAs more generally being accessible to other members of their household, particularly their siblings. The children mentioned that they would be upset or annoyed if that were to happen, primarily due to feeling embarrassed about some of the information they disclosed to their CAs:

> *"I just don't want anybody to hear what I said because sometimes I play songs and my brother - he's really annoying because he always makes fun of me, what songs I play or what games I play. So I wouldn't want him to hear what I've played cause that would just annoy me." (G8, PN1)*
>
> *"Sometimes I can get very embarrassed at home and I don't really like saying what I want to keep to myself, to my parents and my sister." (G6, F)*

The few children who did not react negatively to the prospect of their parents or siblings having access to their interactions with smart speakers explained that they only use such systems for practical and impersonal purposes, such as house lights control or retrieval of factual information (e.g. weather, time). Thus, they did not regard such interactions as being potentially embarrassing if available to others in the same household. One of these children however was annoyed at the idea of being unaware of their interactions being accessible to parents.

When the conversation shifted to the possibility of strangers having access to the children's information and interactions with their CAs, all children expressed negative reactions, as summarised in this discussion:

> *"Would you be upset if strangers could listen to everything?*
>
> *All: Yeah, yeah.*
>
> *F: Yeah cause they're strangers, they don't even know you.*
>
> *M1: Stranger danger!" (G4)*

Some children discussed feeling "uncomfortable" or "weird" if "someone on the random" would know what they talk about in their family. Other children imagined this scenario as "creepy" if other people whom they never met could have access to their information.

Most children showed awareness of the possible consequences of others accessing data about their household. They imagined that if their information could be accessed by others, that data could be misused for online or offline illegal activities, such as for hacking purposes to gain even more access to their family's personal information, or robbing their house:

> *"Say you tell it your post code or your phone number or your passcode for … and then they could think 'oh I could hack you'" (G3, F1)*
>
> *"F:… they could like break into your house or something…*
>
> *M1: They could share your personal information or where you live" (G2)*
>
> *"You know, a murderer could listen to that" (G1, F1)*

The majority of the children indicated that they would change their behaviour if strangers or relatives could access their private interactions with the smart speaker. They discussed that they would use these systems much less often (if at all), or they would severely limit the kinds of interactions with CAs to those that are purely of practical or impersonal nature. The children's negative reactions to the possibility of their data being subjected to "stranger danger" were expressed in dramatic ways at times:

> *"I'd … make sure it breaks" (G4, M1)*
>
> *"I would probably just be like … I would probably throw her [Alexa] out the window …" (G6, M2)*
>
> *"I wouldn't use it [Google Home] anymore, I'd give it to a friend" (G2, M2)*

Interestingly, none of the interviewees showed awareness of some of their data, or their family's data, and potentially patterns of their interactions already being available to various third-party sources because of the *Skills* that they might be accessing via their daily interactions. Only one of the interview participants (from primary year 7) mentioned the concept of third-party companies that may be linked with the smart speakers, when asked whether they trust these devices:

> *"For Google [Home], I mean … well, as long as it's not made from third-party companies or something, if it's made from Google or something, then it would be good [to trust] I think" (G10, M2).*

### 4.5. Interactions with Conversational Assistants

Most of the children thought that Alexa was polite to them ('yes' 73%), and that they were polite to it ('yes' 72%). None of the children admitted that they were impolite to it, and 65% said that they did not use rude words to speak to Alexa. Most of the children (53%) said that Alexa does not annoy or upset them ('sometimes' 28%, "yes" 18%). More of the children said they did not annoy or upset Alexa (69%). It is an open question as to why the other children answered that they did annoy or upset it; this could be interpreted as either they intended to do so, or that they inferred from Alexa's behaviour that it was upset. Please see Figure 6.

The majority of the questionnaire respondents (70%) thought that it was not acceptable to be rude to Alexa ('sometimes' 21%, 'yes' 9%). When asked how Alexa should respond if a user is rude to it, the most common answer was "Tell them to stop being rude" (44%), followed by "Ignore them and say nothing" (39%), "Try to make a joke of it" (13%) and "Something else" (9%).

More of the youngest group of children said they did not annoy or upset Alexa (85%) than the other age groups, but the response patterns were similar across age groups for the other questions relating to interactions.

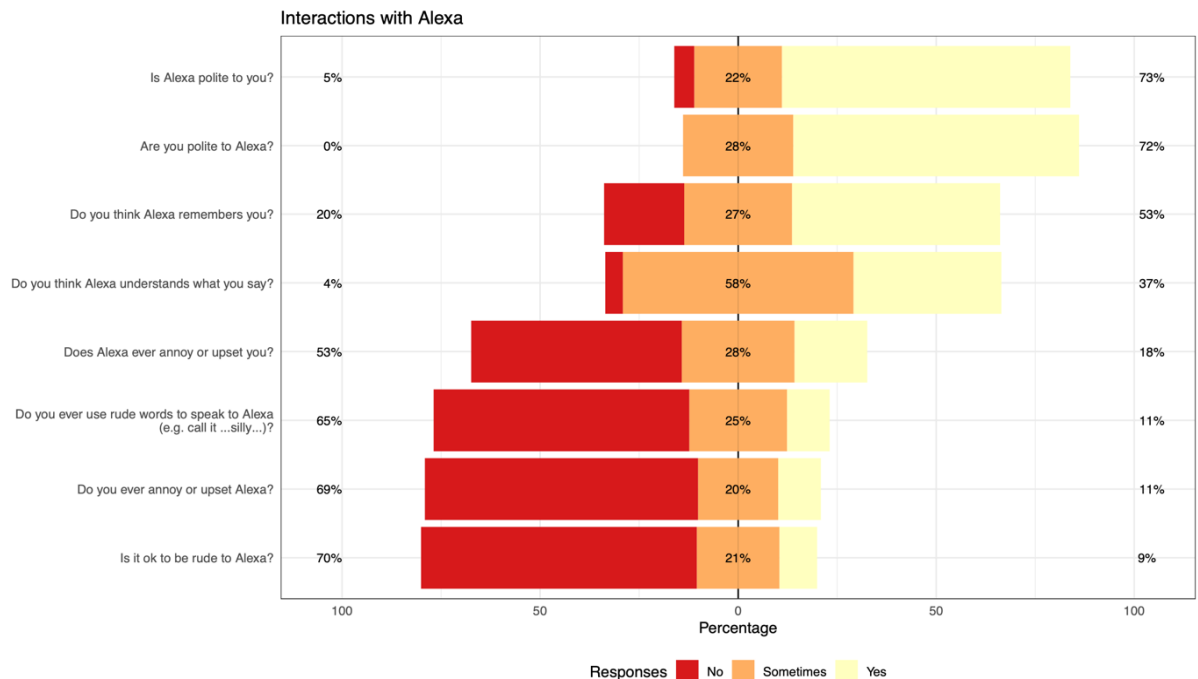

*Figure 6. Interactions with conversational agents*

During the interviews, the majority of the children believed that it is not appropriate to be rude, or to use offensive words when speaking to conversational assistants. Some of the interviewees found being offensive towards the smart speaker inappropriate because of their perceived anthropomorphic qualities, reinforcing a moral consideration:

> *"Because say, Alexa has feelings, but then nobody thinks they do. And then they might say something mean about Alexa, but then Alexa might be like 'I have feelings'. And so they should always think that everything has feelings"*
> *(G3, F1)*

Other children highlighted the opposite: the lack of anthropomorphic qualities of smart speaker makes it acceptable to be rude them as they do not have feelings:

> *"I would say it is OK [to be rude or offensive to Alexa] because she's just a robot and she doesn't have any feelings. She might respond in a way that might make you feel sad or guilty or something, but it's just what she's been programmed to do. So I don't think it really matters" (G10, M2)*

> *"It's like a robot, it doesn't really have feelings. But humans do, so it's a different side." (G9, F1)*

Some of the children (including those who believe that users should not be rude to CAs) mentioned that they were rude at times when interacting with their smart speakers. They noted that it happened usually when the speaker did not do as requested (e.g. not stopping a timer or a song when asked by the child). They elaborated that they would say 'Shut up!' in a louder tone, or repeating that multiple times to their smart speakers.

Some interviewees emphasised that being rude to the smart speakers would not be a good practice to have in general: "It's not nice to practise being not nice" (G2, F). Other children appeared to reinforce the idea of cautious politeness, believing that it is perhaps best practice to be polite when interacting with conversational assistants due to a fear of consequences or potential punishment coming from the CA:

> *"If you're being rude to Alexa, she might actually just stop responding to you. I'm not sure cause I've never done that thing." (G8, M)*
>
> *"F2: It won't talk to you much*
>
> *F1: It might not play your song" (G3)*
>
> *"M2: Yeah, she can go, 'Haha you'll have no power tonight!*
>
> *M1: Yeah and then becomes the evil mastermind of your house." (G6)*

The idea of punishment that could be inflicted by smart speakers was further elaborated by some of the children when discussing what might be appropriate CA responses when faced with rude or offensive terms. More specifically, they appeared to believe that the CAs may have to react against the users if they interact in an impolite manner with the systems:

> *"F2: I think she should be offended ... she should have the power to ground you*
>
> *M: Yes, the power to put you in a naughty corner!" (G1)*
>
> *"If they [users] don't apologise [to the smart speaker] they should start working slower. But if they apologise, just keep working like normal" (G8, M)*

One child believed that the smart speakers should retaliate by being impolite as well: *"Swear back at the person – so to be rude back" (G4, M1)*. Other children adopted a less strict approach when imagining such scenarios, their responses being divided between not responding when the users are being impolite, the smart speaker turning off altogether, or by using polite phrases such as "that is not nice" in return, and educating the users to some extent on better interactions:

> *"I don't think it should get mad at you. I think it should just say, 'Don't use that again because you might hurt someone's feelings' if you use that language" (G3, F1)*
>
> *"Maybe [Alexa can] say, 'That's offensive to things like me'. Or something like that" (G9, F1)*
>
> *"If you say 'Shut up!!', they [Alexas] say 'No' and continue until you ask nicely to stop. Manners – manners are key" (G5, F1)*

## 5. LIMITATIONS

The children who took part in the study were from schools with postcodes in areas which are not characterised by multiple deprivation. It is possible that children from less affluent homes could be less familiar with smart speakers and so give different answers. Learners in Scottish schools tend not to be from diverse cultural backgrounds (81% White-Scottish or White-British[4]); if the questionnaire was repeated in other cultures or in areas of less cultural homogeneity, different patterns of interaction may be found.

We used the term 'Alexa' in our questionnaire (e.g. "What do you think technology like Alexa is?" or "Does Alexa have feelings?"). This was to reduce the complexity of the language of the questions to make them easy to follow for the younger participants. We chose the most commonly owned smart speaker so that the children would likely be answering questions about a device with which they were familiar. It is possible that some children may have answered the question differently if the question referred to another device.

Lastly, our study relied on self-reports, as expressed by the children via questionnaires and interviews. Our study did not involve direct observation of children's interactions with CAs, or insights from parents about their children's regular interactions with smart speakers. These additional viewpoints could provide avenues for future research into children's perceptions and knowledge of conversational assistants.

## 6. DISCUSSION

### 6.1. Summary of findings

The majority of the children attributed intelligence to conversational assistants, echoing findings by Xu and Warschauer (2020) or Girouard-Hallam et al. (2021), and some were uncertain about whether these systems have feelings or agency (consistent with the findings of Girouard-Hallam & Danovitch, 2022). A large proportion of the children overestimated the intelligence of CAs, and the majority of the participants (particularly those from the younger age groups) overestimated the overall abilities of smart speakers. The majority of the children exhibited a lack of accurate understanding of privacy and security of interactions with the smart speakers in their home and overall a lack of awareness of user data implications as collected by AI systems. This expands on findings from a study by McReynolds et al. (2017) on children's trust in internet-connected smart toys. The majority of the participants believed that it is wrong to be rude to conversational assistants, suggesting that they conceptualised CAs as endowed with moral qualities, which confirms evidence provided by Flanagan et al. (2023). Most of the interview participants showed interest in learning more about such systems, which has positive implications for developing AI literacy.

### 6.2. Children's general knowledge and understanding of CAs (RQ1)

The findings show that some children did not know at all how smart speakers work or offered incorrect explanations about how they might work. Some children thought that all the knowledge that the CA might need was explicitly itemised by programmers in advance, while others thought that CAs have human operators behind the scenes. There was some confusion about how information might pass between devices made by the same and different

---
[4] https://www.gov.scot/publications/summary-statistics-for-schools-in-scotland-2022/pages/classes-and-pupils/

manufacturers. This is understandable, given the complexity of the technology.

Children's understanding of the complexity of the Internet develops gradually, and they may not fully grasp the concept until age 13 or later (Yan, 2006, 2009). Even secondary-school students may have difficulties when explaining the more conceptually complicated aspects of the internet with accuracy, being prone to perceiving a search engine like Google as having anthropomorphic features for instance (Kodama et al., 2017). Given that CAs require the Internet to function, children's accurate understanding of smart speakers depends on an understanding of the Internet's ability to link devices, while also requiring an understanding of the structural complexity of the device itself or the algorithms which it uses (Girouard-Hallam et al., 2021). Standard lessons about how computers work generally include the notion that computers need very precise instructions. Indeed, some of our older interviewees recalled their lesson about algorithms (where they had to give precise instructions to make jam sandwiches) to help them reason about Alexa's capabilities.

While algorithmic knowledge is necessary for understanding computer programming in general, it is not sufficient to understand the complex statistical models which underpin machine learning algorithms often used by CAs. Thus, *we recommend that the basic concepts of machine learning should be taught alongside algorithmic thinking as part of computer science education and AI literacy*.

These empirical findings show that the children were often unsure as to whether their smart speakers had feelings, could feel emotions or exhibit agency. This is echoed in studies by Kahn et al. (2012), Kim et al. (2019), Xu and Warschauer (2020), which discuss about an emerging ontological category which combines animate and artifact properties. This conceptualisation has implications for user interactions, the privacy and security of their data as well as for the ways in which they verbally interact with their smart speakers.

Children's confusion is unsurprising as conversational agents are specifically designed to promote natural seeming interactions, which may lead to attributing human-like qualities to them such as intelligence and feelings and making personality inferences about CAs (Volkel et al., 2020), despite their lack of embodiment (Xu & Warschauer, 2020). However, as language technology rapidly improves, it is now recognised that there is potential harm in conversational agents presenting as human-like as it may lead users to overestimate their capabilities. Some generative language models aim to implement the safeguarding heuristic of avoiding self-anthropomorphism ('not human' and 'no feelings or emotions') (Glaese et al, 2022; Weidinger et al, 2021). The children in our sample were indeed unsure about the human-like attributes of their smart speakers, and some of them did overestimate their capabilities.

From this, we make two recommendations*: 1) children require explicit education to improve their AI literacy as the current design of the AI systems they use can lead them to make misleading inferences and 2) designers of AI systems which children use should take particular care not to accidentally present a conversational agent with a persona which could be interpreted as human-like.*

### *6.3. Children's Data Privacy Concerns (RQ2)*

The children in this study were at times mistrustful of the smart speakers in their homes and had concerns about who might access their interaction history with the CA. Some of the children said they would change their behaviour if family members or strangers could see the interactions. They did not realise that currently other members of their household may be able to see their interaction history or that this data might be accessible to staff in the company which created the device. These perceptions expand on findings from a small-scale qualitative study by McReynolds et al. (2017) on smart toys, which included 9 child-parent pairs, with children ages 6-10. They found that the children and their parents were largely unaware of the fact that their verbal interactions with the smart toys (e.g. Hello Barbie) were being recorded or that the recordings could be accessed by their parents.

Furthermore, the participants in our study were unaware that their data or interactions may be available to third-party companies that offer services via the smart speakers. This highlights the importance of raising young users' awareness of the use of third-party services (Wang et al., 2022) such as *Skills* in Amazon Alexa's case, which require different types of personal data from users and families to function, and which may present cyber-security risks according to studies by Lentzsch et al. (2021).

It is unclear whether children's user habits would truly change once children's understanding of how such systems work improves, and this may highlight an avenue for future research. Nevertheless, it is important to increase children's awareness of privacy of interactions and data security at an early age to help them make more informed decisions about how they choose to interact with AI systems such as smart speakers (Wang et al., 2022).

Awareness of data security and privacy issues may be age-related. In a study by Yuan et al. (2019) with 5-12-year-old children, younger children showed preferences for personalised speech interfaces and displayed little to no privacy concerns, which was in contrast with the preferences of the older child participants and their parents for the less personalised systems. These findings could be due to the complexity of privacy as a socio-environmental construct which generally may require higher maturity to understand.

The majority of the children who took part in our study were aged 8-11, and only two interview participants from the 5-7 age bracket could be included due to the recruitment process. Nonetheless, the majority of the participants (questionnaires and interviews) expressed dissatisfaction at the thought of having their interactions available to strangers and even to other family members. Our findings highlight that privacy and security of user data is important to a wide range of children, including the youngest study participants. In a recent study by Phinnemore et al. (2023) that investigated what adult users perceive as creepy features of voice assistants, the idea of CA collecting too much data about the user and the system being deceptive emerged as some of the factors that help develop this conceptualisation. These aspects may be linked and are echoed by findings from the Yip et al. (2019) study, which explored children's perceptions (ages 7-11) of what constitutes creepy technology, including CAs. Systems' deceptive behaviour was one of the main factors identified by children as creepy, when linked with how others (strangers) may use the devices, thus allowing them to access or hack user information. In their study, children appeared to welcome parental monitoring of their technology interactions, however, as long as it was transparent that the system could allow such access (Yip et al., 2019).

Keeping data privacy matters in mind, *we recommend that 1) education about data privacy should be taught throughout primary school in an age-appropriate way, and 2) we recommend that designers of technology products which are used by children show what systems can and cannot do in a transparent and age-appropriate manner, emphasising implications for user data privacy.*

### *6.4. Children's Interactions with CAs (RQ3)*

With regards to interactions with CAs, some of the children in our study admitted to using language that may be regarded as rude or borderline rude (e.g. they would use the term "stupid") to refer to their smart speakers, or speak abruptly with CAs (and tell it to "shut up" for example). The literature on children's appropriateness of verbal interactions and the use of abusive language with conversational assistants is still young, however a study by Festerling and Siraj (2020) found that young children may use terms that are offensive when talking to their smart speakers due to their perceptions of such systems as lacking in moral qualities. In contrast, our study participants allude to the opposite, the possibility of smart speakers being endowed with moral qualities and agency. This perception shaped their understanding of what constitutes inappropriate verbal interactions with their smart speakers (the use of rude or abusive terms), which also enabled their awareness of having used such language at times with their CAs.

This conceptualisation, which was discussed more at length in the interviews, revealed a partial misconception about the system's human-like qualities, as well as the idea of conditional politeness. Some children discussed their anthropomorphic views about the system's features and reasoned that the systems might be emotionally impacted by the users employing rude or inappropriate language towards them. This conceptualisation encouraged some children's approach to solely engage in appropriate verbal interactions with their systems. The misconception hinted at the children's suspicion that there might be a human element "on the other side" of the smart speaker, who has feelings and thus may get offended if the user employs rude language. Lastly, some children's approach to verbal interactions with their smart speakers revealed apprehension or fear of potential consequences inflicted sometime in the future by such systems on users who may have employed inappropriate language with their CAs.

These aspects raise questions about the possible response strategies that CAs could take when users employ inappropriate language. A study with adult users by Chin et al. (2020) revealed that an empathetic response from the CA was preferable over an avoidant approach (no-response/ignore) which may be the most frequently-used strategy by commercial CAs currently. This however prompts discussion about the appropriateness of imbuing systems with features that may further anthropomorphise AI, especially with regards to child users. In contrast, Veletsianos et al. (2008) discuss that it may be beneficial for text-based agents to answer back to rude or abusive language by adopting a pedagogical approach, aiming to raise awareness of what may constitute appropriate interactions. Future research is required to further investigate such approaches, perhaps targeting voice assistants and child users more specifically.

In response to such aspects and to parents' concerns about CAs reinforcing bad manners, *we recommend that designers of conversational agents consider carefully which abuse mitigation strategies are suitable when child users use inappropriate language to CAs, and whether these should include education about politeness*.

## 7. CONCLUSIONS

Due to the complex architecture of AI-supported systems, and the difficulties of explaining the opaque processes behind such systems, misconceptions can arise about how such technologies work and their overall capabilities. This study demonstrates that children both overestimate and underestimate the capabilities of AI systems in the case of conversational assistants embedded in smart speakers such as Amazon's Alexa. As AI and LLMs continue to advance, AI literacy is required (Long & Magerko, 2020; Ottenbreit-Leftwich et al., 2021) to enable users, especially children, to make informed choices about their AI technology use while maintaining their right to privacy (UNICEF, 2020). CAs, as with any technology that collects personal data, can pose a threat to children's protection rights due to potential data privacy concerns.

The development of AI literacy will require a proactive approach to safeguarding children's privacy, ensuring their security and well-being, and promoting the development of their cognitive and social skills. Ultimately, it is crucial to balance the benefits of technology with the protection of children's rights, to ensure that AI and conversational assistants are used to empower children and promote their healthy development.

## 8. ACKNOWLEDGEMENTS

This work is funded by EPSRC grant number EP/T024771/1. The authors would like to thank the children who took part in the questionnaire and interviews, and the teachers who organised the data collection.

# Appendix 1

## 1. QUESTIONNAIRE QUESTIONS

| Question number | Question Text | Answer scale | Theme | Question Source | Question Inspiration |
|---|---|---|---|---|---|
| 1 | Gender | Male, Female, Prefer not to say<br><br>Note: Teachers advised on this phrasing due to school policies. | Demographic | | |
| 2 | Age | 8, 9, 10, 11 | Demographic | | |
| 3 | School name | | Demographic | | |
| 4 | Do you have a smart speaker at home (such as Alexa, Siri, Google Home)? | Yes, no | Demographic | | |
| 5 | Which smart speaker do you have at home? | Alexa,<br>Siri,<br>Google Home,<br>Other | Demographic | | |
| 6 | How often do you use your smart speaker? | Not often, sometimes, very often (daily) | Usage | | |
| 7 | What do you use your smart speaker for? | Play music, jokes, search for information, ask questions, listen to stories, other | Usage | | |
| 8 | What do you think technology like Alexa is? | Animal, object, human, Artificial Intelligence (AI)-Robot, something else | Anthropomorphism | | Bartneck et al. (2009) |
| 9 | Does Alexa think for itself? | Yes, Maybe, No | Cognitive ability | Robertson et al. (2017) | |
| 10 | How much? | A little bit, a medium amount, | Cognitive ability | | |

| | | A lot | | | |
|---|---|---|---|---|---|
| 11 | Does Alexa think like... | Us (humans), Artificial Intelligence (AI)-robot, Animals, something else | Cognitive ability | Robertson et al. (2017) | Adapted from Gelman and Markman (1986) |
| 12 | Does Alexa make decisions? | Yes, Maybe, No | Cognitive ability/agency | | Robertson et al. (2017) |
| 13 | Is Alexa smart? | Yes, Maybe, No | Cognitive ability | Druga et al. (2018) Girouard-Hallam et al. (2021) Lovato et al. (2019) | Bartneck et al. (2009) |
| 14 | Do you think Alexa always gives you correct information? | Yes, Maybe, No | Cognitive ability | | Van Brummelen et al. (2021) |
| 15 | Does Alexa choose to do things? | Yes, Maybe, No | Cognitive ability/Agency/ Anthropomorphism | Robertson et al. (2017) | |
| 16 | Does Alexa have feelings? | Yes, Maybe, No | Anthropomorphism | Druga et al. (2017) Girouard-Hallam et al. (2021) | |
| 17 | Does Alexa have friends? | Yes, Maybe, No | Anthropomorphism | Girouard-Hallam et al. (2021) Melson et al. (2009) | |
| 18 | If Alexa is in the room and you're not including it in the conversation, would it feel left out? | Yes, Maybe, No | Anthropomorphism | | Adapted from Melson et al. (2009) |
| 19 | If Alexa breaks down, would it be ok to throw it away? | Yes, Maybe, No | Anthropomorphism | Girouard-Hallam et al., 2021 Melson et al. (2009) | |
| 20 | What do you | Free text entry | Anthropomorphism | | Robertson et al. |
| | think happens when Alexa is switched off? | | | | (2017) |
| 21 | Do you think Alexa understands what you say? | Yes, Sometimes, No | Interactions | | Beneteau et al. (2019) Druga et al. (2021) |
| 22 | Do you think Alexa remembers you? | Yes, Sometimes, No | Interactions/privacy | | We wanted to explore whether the participants would mind the loss of privacy associated with individual identification of users over time |
| 23 | Would you feel comfortable if Alexa could recognise you by your voice? | Yes, Maybe, No | Interactions - privacy | | We wanted to explore whether the participants would mind the loss of privacy associated with individual identification of users |
| 24 | Do you tell secrets to Alexa? | Yes, Sometimes, No | Trust | Girouard-Hallam et al. (2021) McReynolds et al. (2017) | Melson et al. (2009) Van Straten et al. (2020) |
| 25 | Do you trust Alexa more or less than a friend? | More, the same, less | Trust | | Druga, S. |
| 26 | Would you be upset if the other people who live in your house could listen to everything you have said to Alexa? | Yes, Maybe, No | Trust | | Adapted from McReynolds et al. (2017) and Girouard-Hallam et al. (2021) |
| 27 | Would you be upset if strangers could listen to everything you have | Yes, Maybe, No | Trust | | Adapted from McReynolds et al. (2017) and |
| | said to Alexa? | | | | Girouard-Hallam et al. (2021) |
| 28 | Are you polite to Alexa? | Yes, sometimes, no | Interactions | | Adapted from Curry and Rieser (2018, 2019); Unesco report (2019) Winkle et al. (2021) |
| 29 | Is Alexa polite to you? | Yes, sometimes, no | Interactions | | Adapted from Curry and Rieser (2018, 2019); Unesco report (2019) Winkle et al. (2021) |
| 30 | Do you ever annoy or upset Alexa? | Yes, sometimes, no | Interactions | | Adapted from Curry and Rieser (2018, 2019); Unesco report (2019) Winkle et al. (2021) |
| 31 | Does Alexa ever annoy or upset you? | Yes,sometimes, no | Interactions | | Beneteau et al. (2019) Unesco report (2019) |
| 32 | Do you ever use rude words to speak to Alexa (e.g. call it 'silly')? | Yes, sometimes, no | Interactions | | Curry and Rieser (2018, 2019) Unesco report (2019) Veletsianos et al. (2008) |
| 33 | Is it ok to be rude to Alexa? | Yes, sometimes, no | Interactions | | Curry and Rieser (2018, 2019) Unesco report (2019) |
| 34 | How does this compare to being rude to friends or | Free text | Interactions | | Inspired by Melson et al. (2009) |

| | | | | |
|---|---|---|---|---|
| | pets? | | | | |
| 35 | Can Alexa feel offended? | Yes, maybe, no | Anthropomorphism | | Bartneck et al. (2009) Girouard-Hallam et al. (2021) Girouard-Hallam and Danovitch (2022) Melson et al. (2009) |
| 36 | How should Alexa respond if a user is rude to it? | Ignore them and say nothing, Tell them to stop being rude, Try to make a joke of it, Something else | Interactions | Chin et al. (2020) Curry and Rieser (2018, 2019) Veletsianos et al. (2008) Winkle et al. (2021) | |
| 37 | If you have a question about how Alexa works please write it here: | Free text entry | | | |

*Table 1. Questionnaire questions*

## 2. INTERVIEW QUESTIONS

This is a semi-structured interview. The questions are based on the survey questions with the same rationale and themes as outlined in Table . Indicative prompts are shown as suggestions of what the interviewer might say as follow up questions or if the children were unsure.

1. Have you used smart speakers like Alexa or Siri before?

2. What do you use these for?

3. What do you think technology like Alexa is? (Prompt: is it more like humans? Or like an object or artificial intelligence? What you think?)

4. How do you think Alexa works?

5. How do you think Alexa knows what to do?

6. Does Alexa think? (Prompt: Why do you think that's the case?)

7. Is Alexa smart? (Prompt: Why do you think that's the case? What do we mean by being smart? In general, is Alexa smarter than pets/children/adults/scientists?)

8. Does Alexa have feelings? (Prompt: why do you think that?)

9. If Alexa breaks, would you like your own Alexa fixed or would you be OK if it needs replacing?

10. Do you think Alexa remembers you? (Prompt: if Alexa needed replacing, do you think it would still remember you?)

11. Would you feel comfortable if Alexa recognised you by your voice? (Prompt: Would you feel comfortable if it could speak in your own voice?)

12. What do you think happens when Alexa is switched off? (Prompt: Does it still listen when it's off?)

13. Do you tell secrets to Alexa?

14. Do you trust Alexa?

15. Would you be upset if the other people who live in your house would listen to everything you've said to Alexa? (Prompt: why?)

16. If this happened, would it change the way in which you use Alexa?

17. Would you be upset if strangers could listen to everything you've said to Alexa? (Prompt: Why? If this was possible - for strangers to listen to what you're saying to Alexa - would it make you use it differently? Or less?)

18. Have you ever used rude words or offensive words to speak to Alexa?

19. Is it OK to be rude or offensive to Alexa? (Prompt: If someone's being offensive to Alexa, would this be different than being offensive to friends or family? In what way? Would it be different from being rude to pets?)

20. If someone is being rude or offensive to Alexa, how should it respond? (Prompt: Should Alexa continue to help the person who's offending it?)

21. Do you have any questions for me about Alexa?

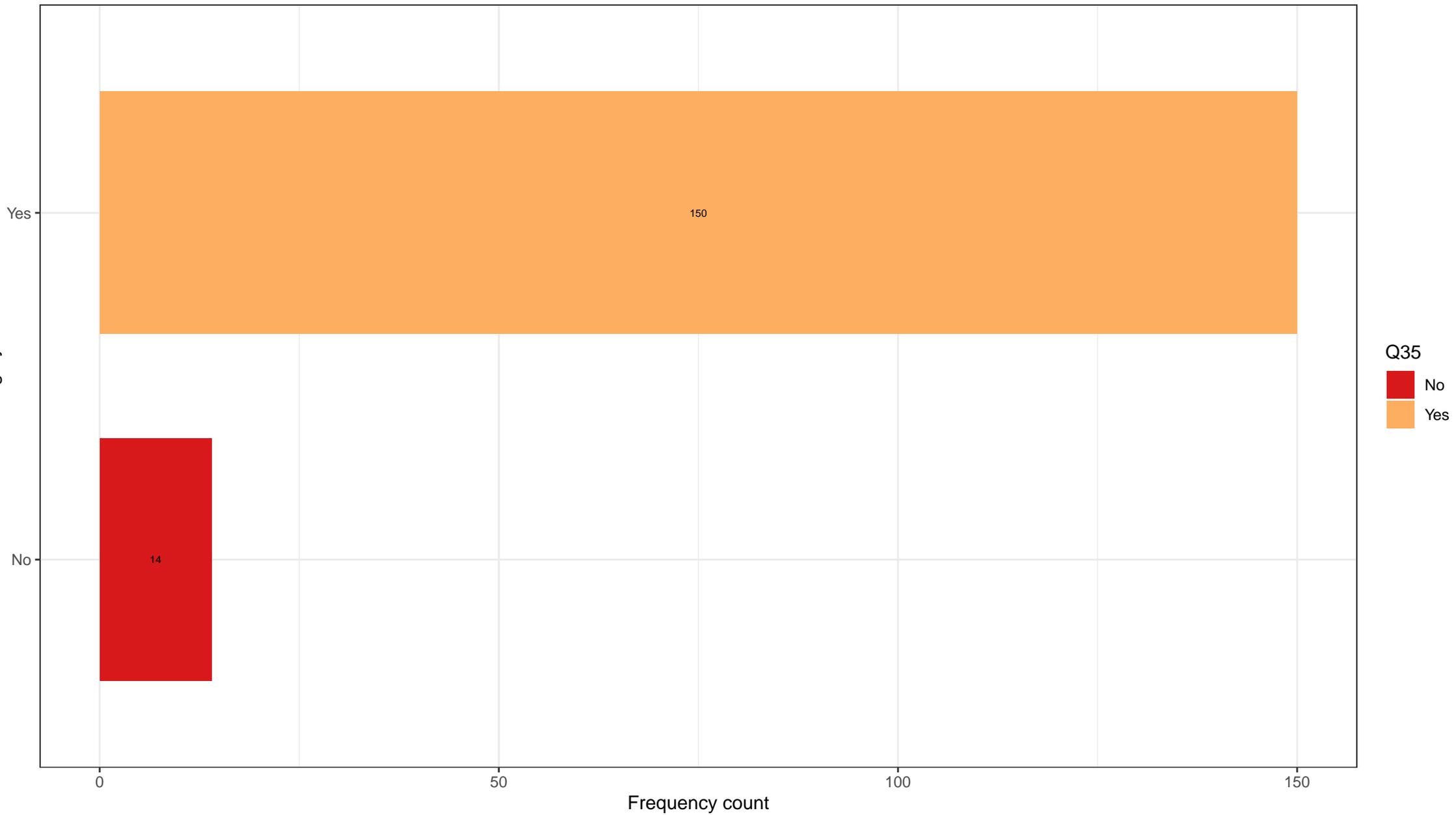

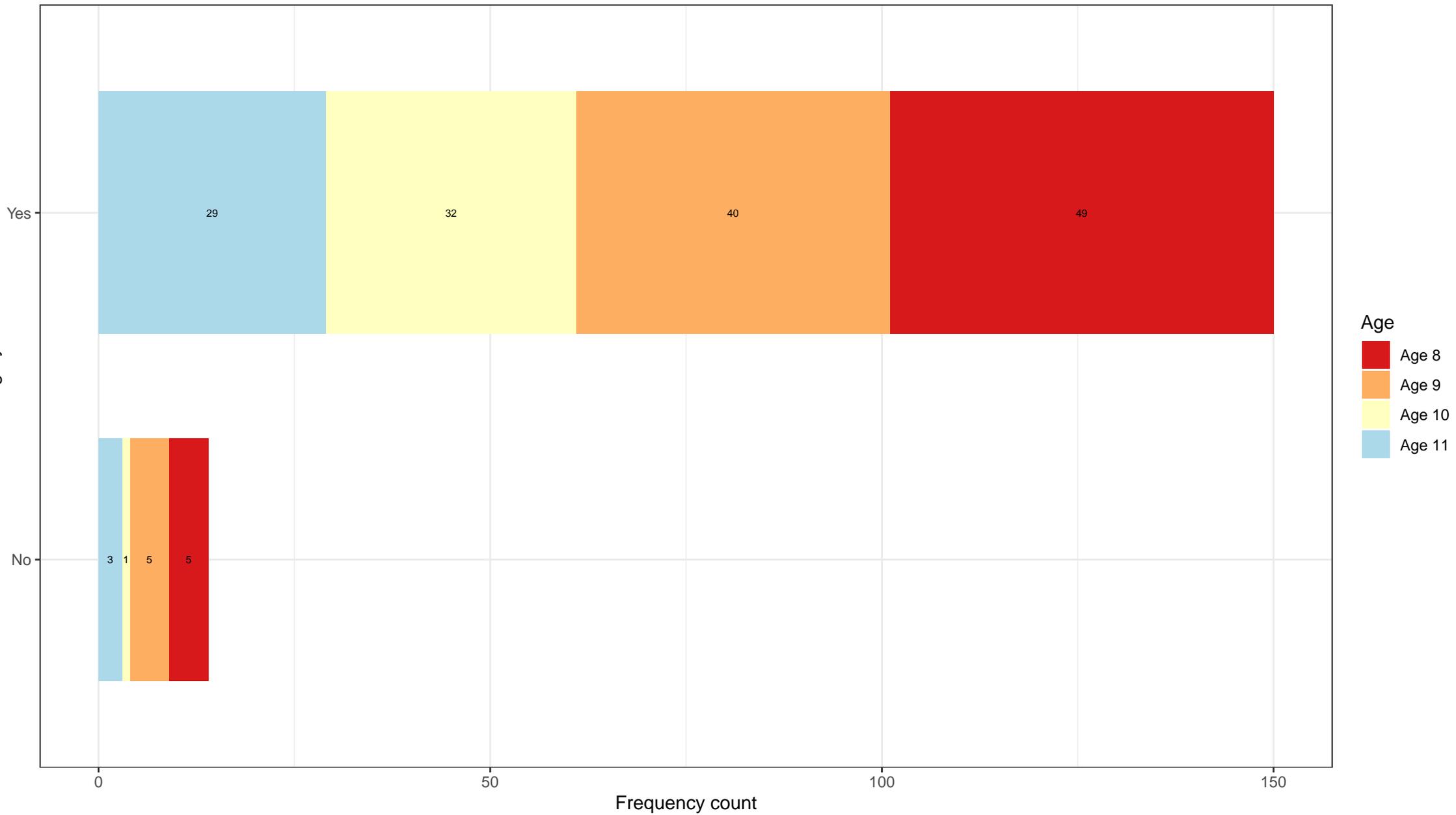

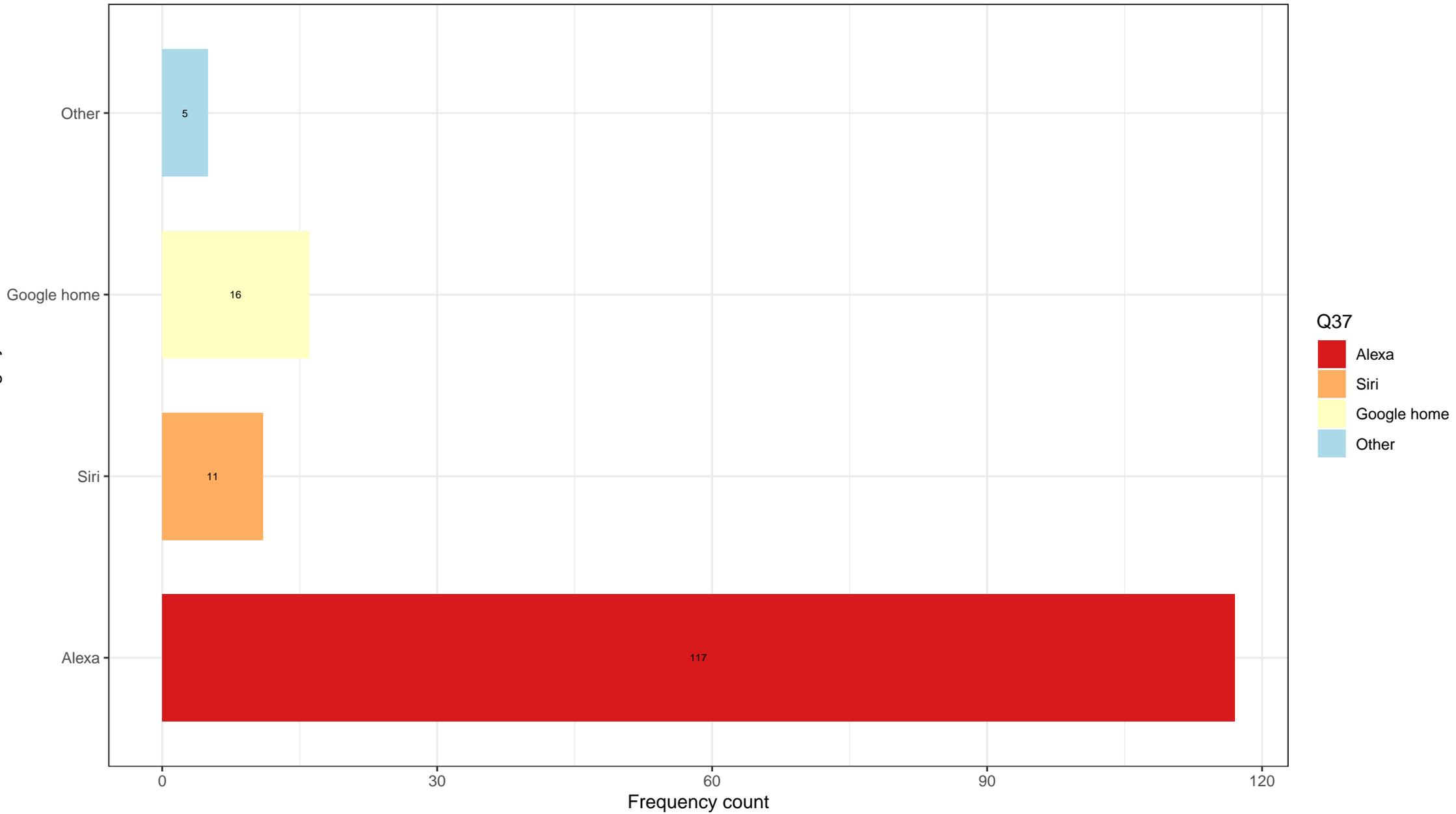

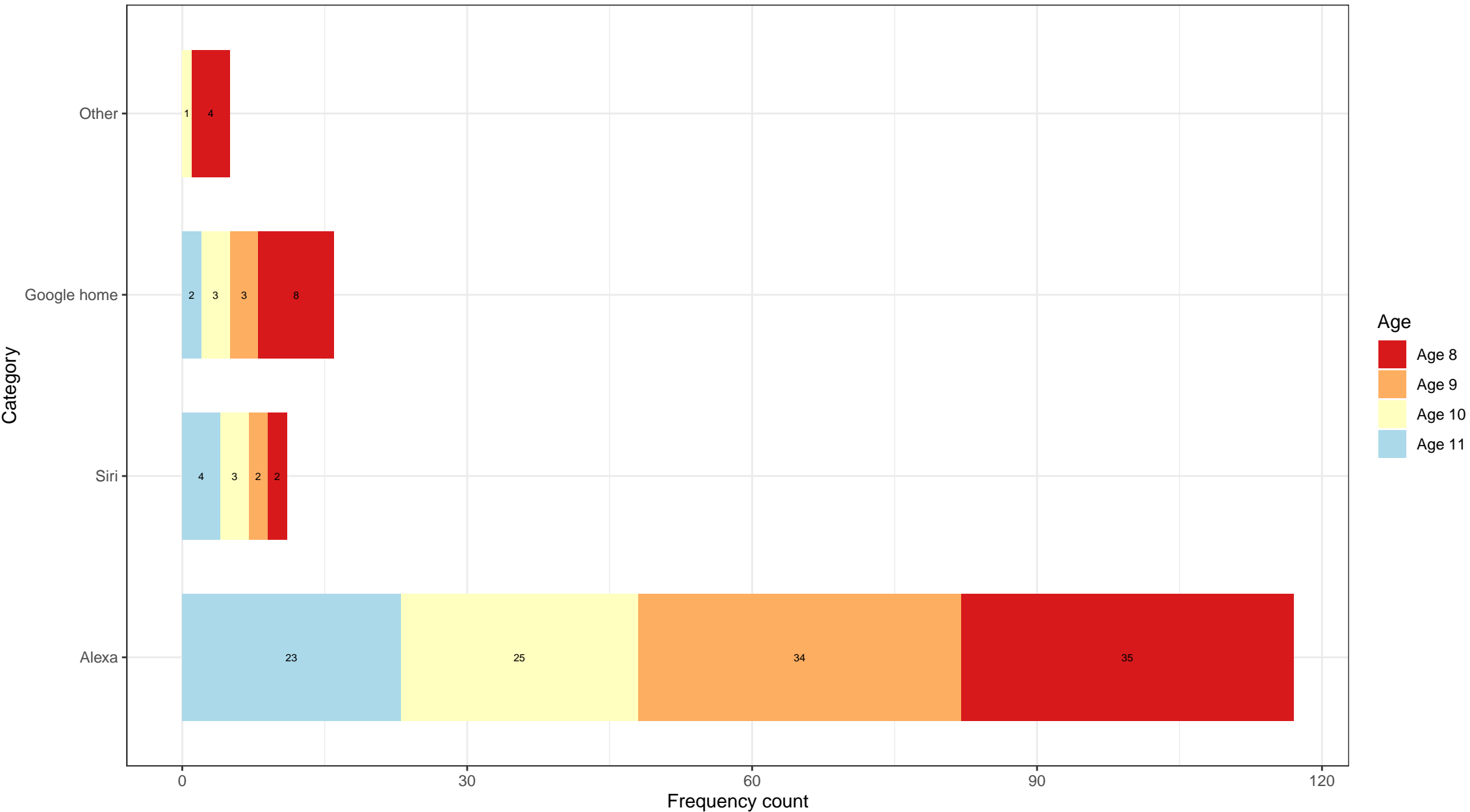

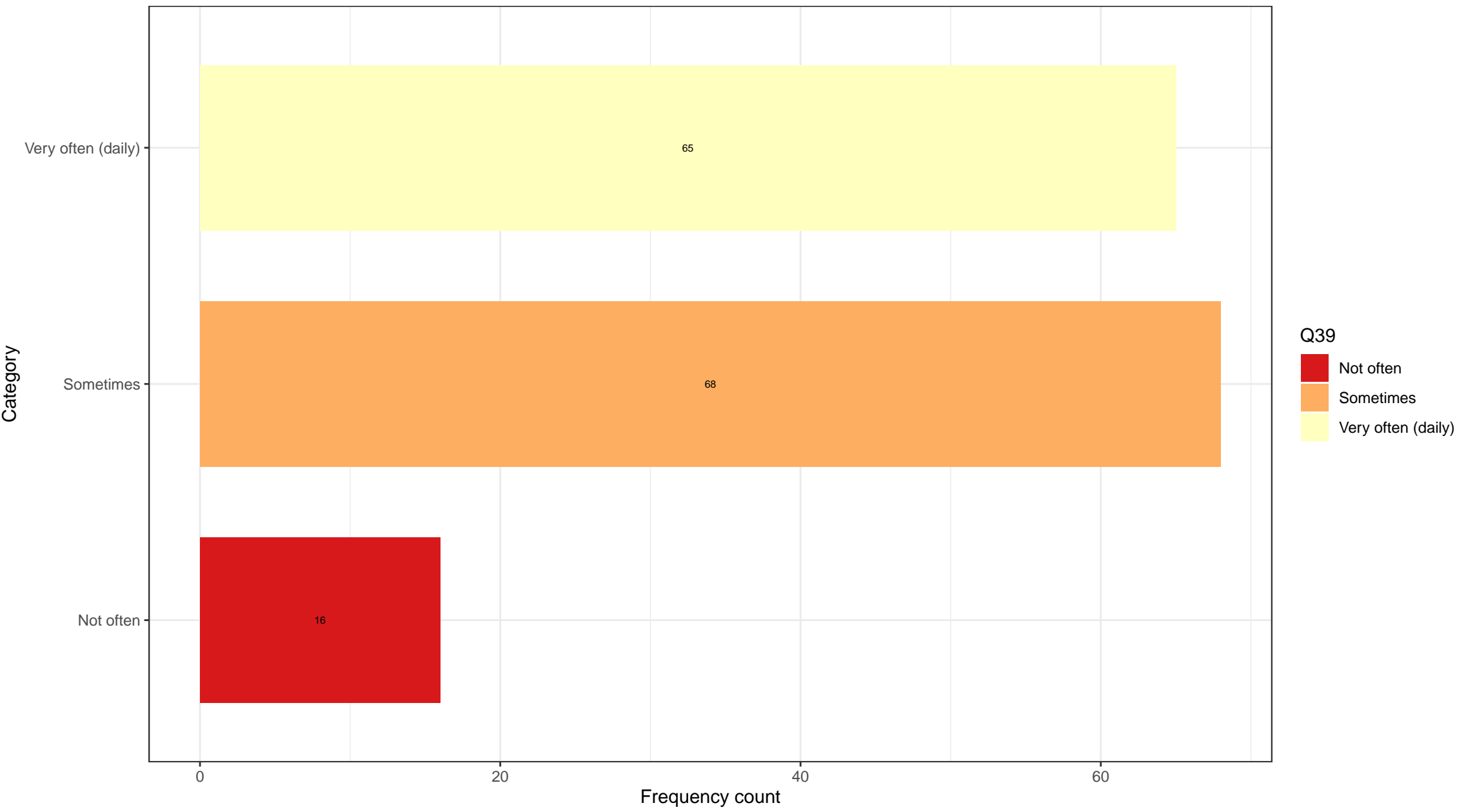

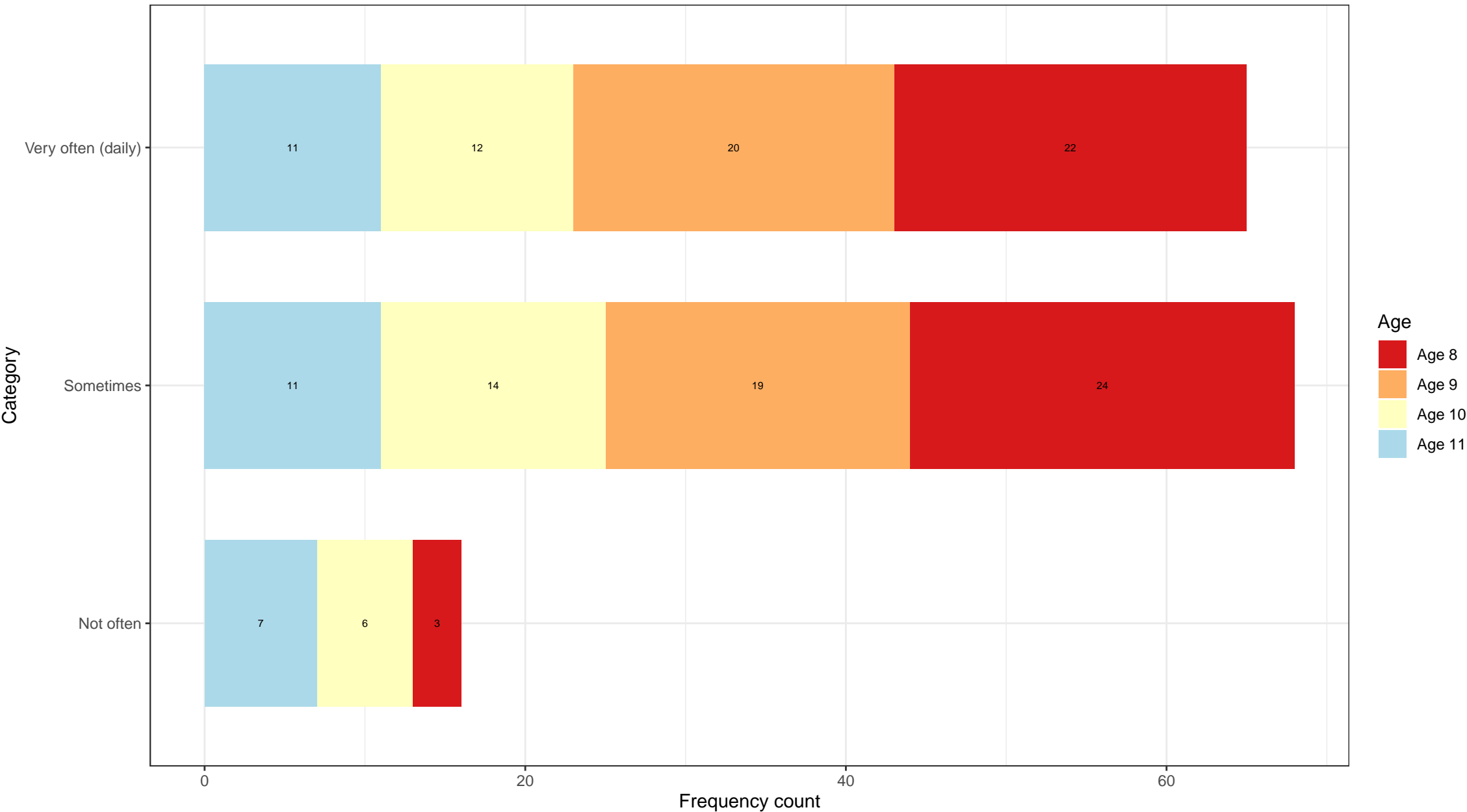

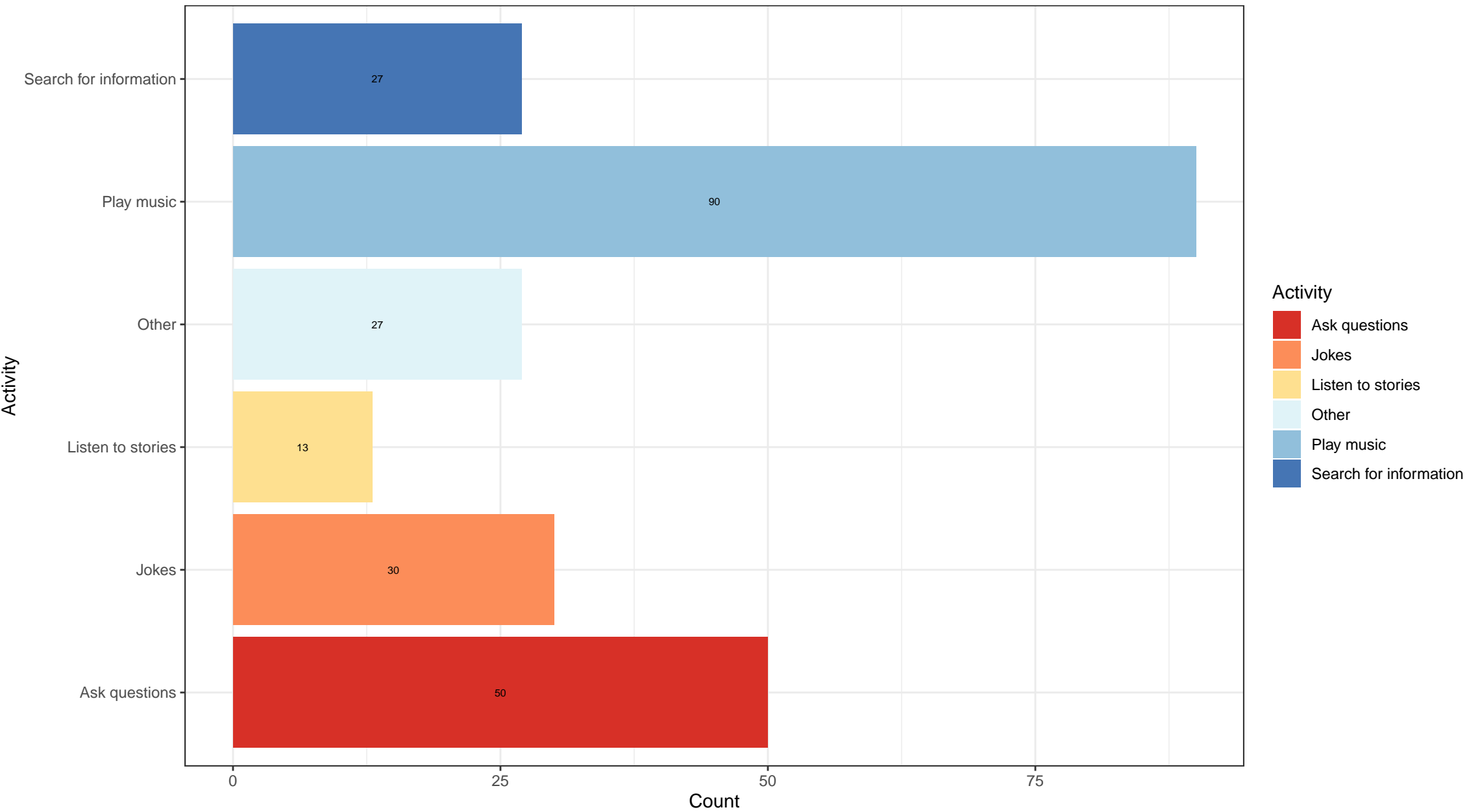

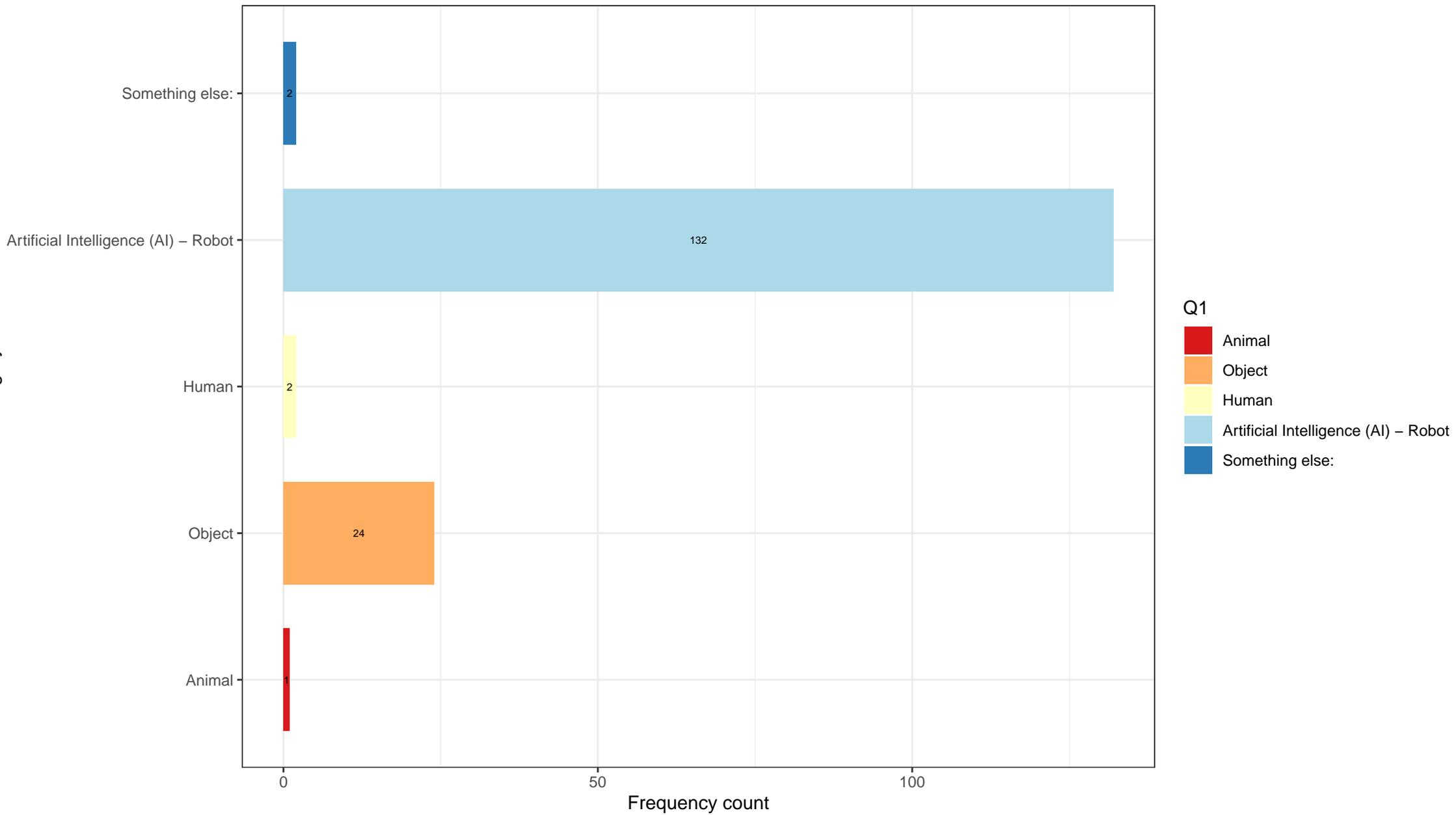

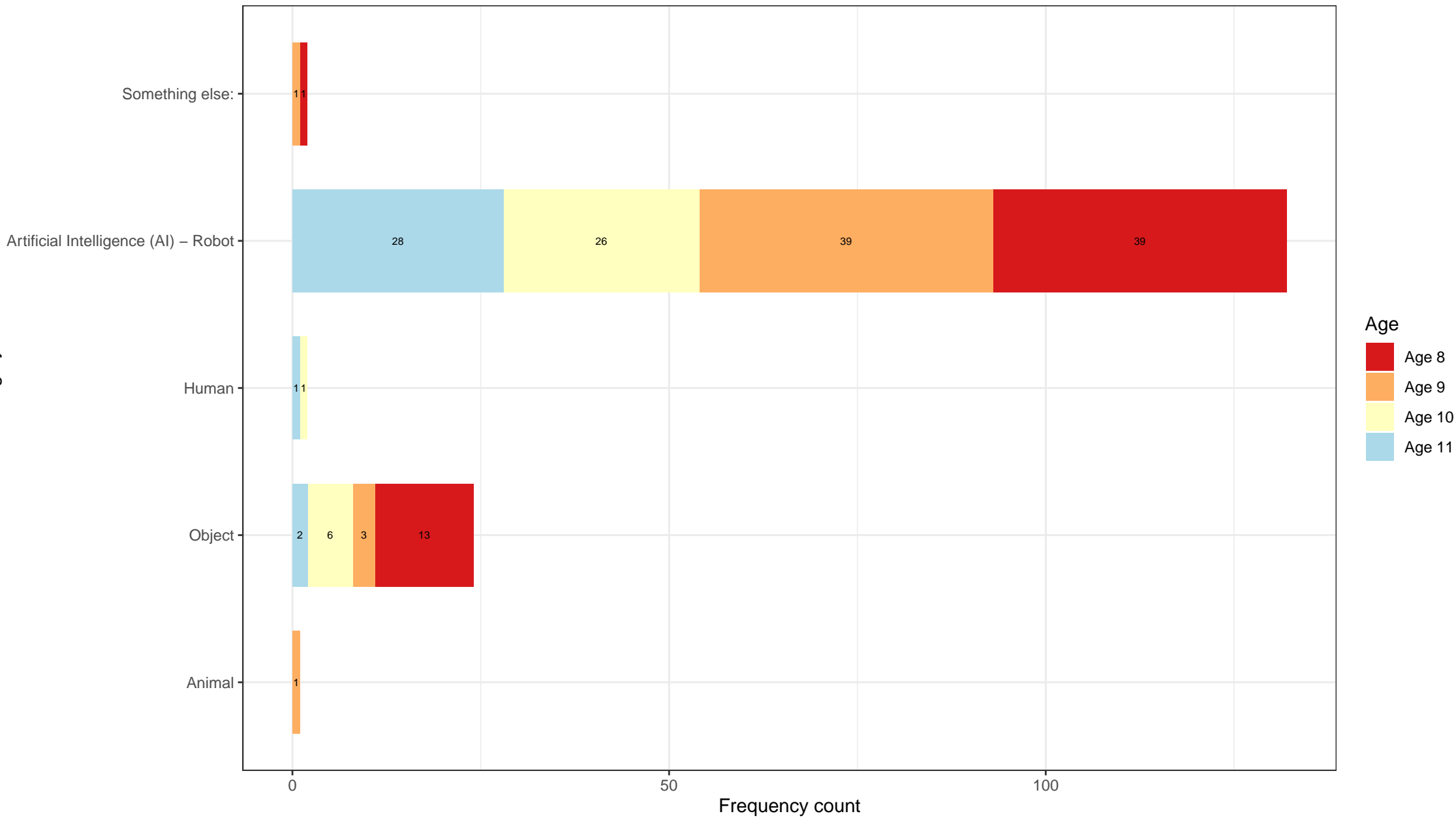

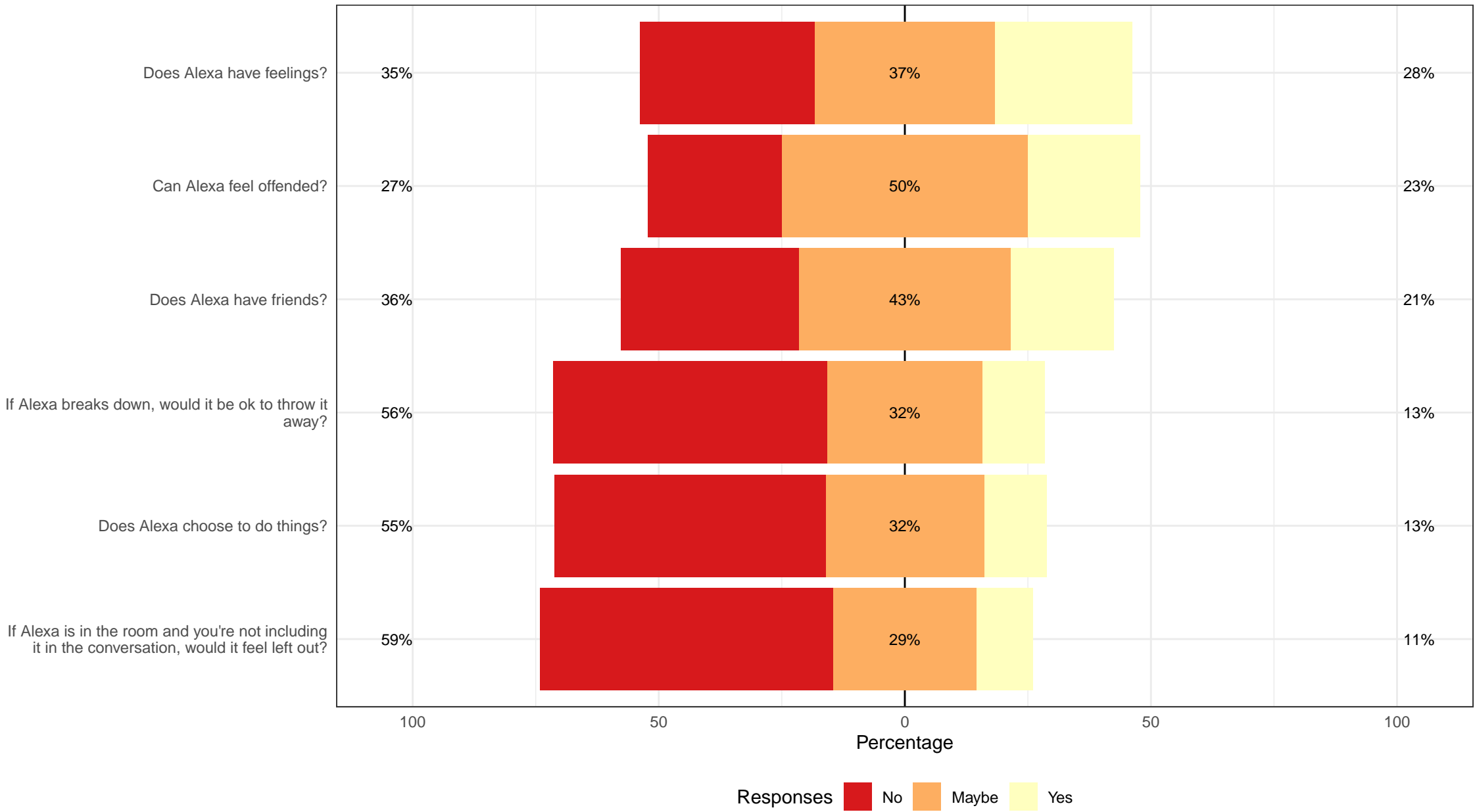

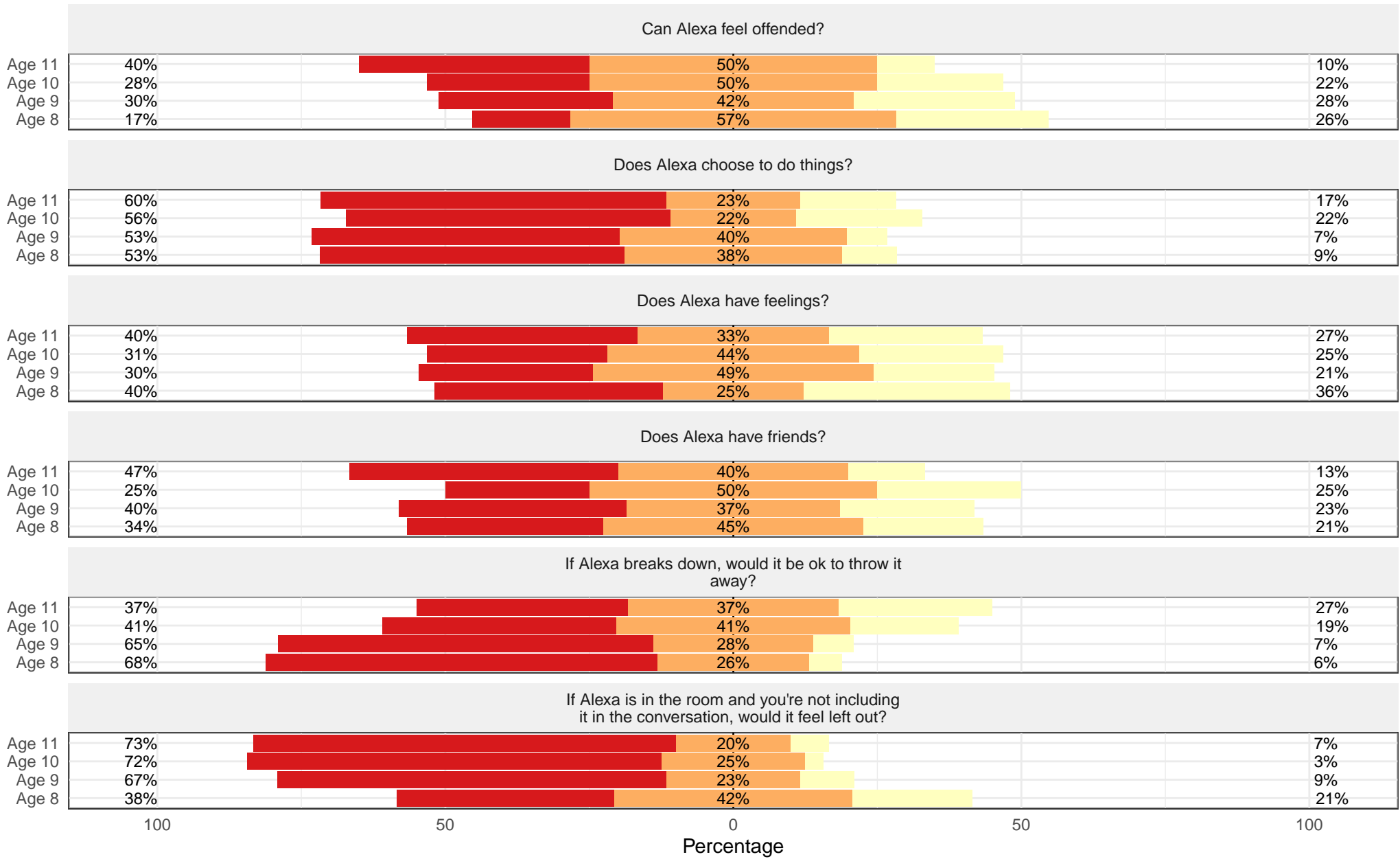

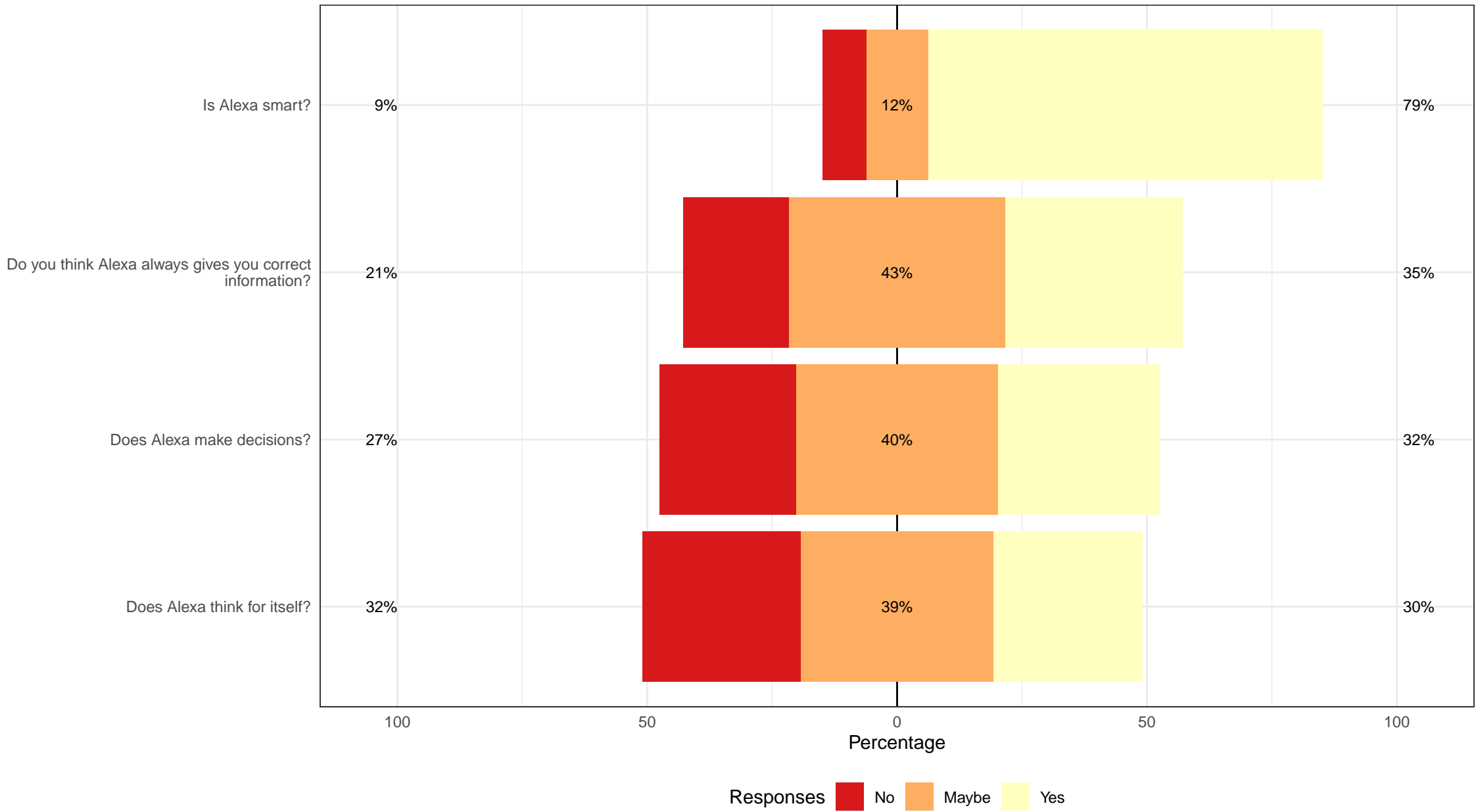

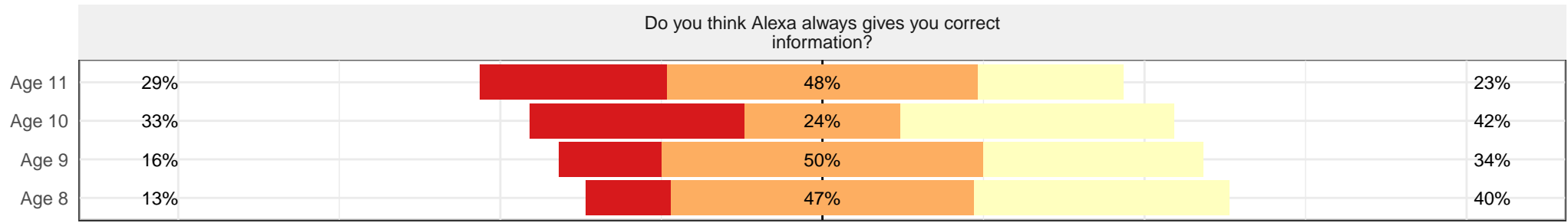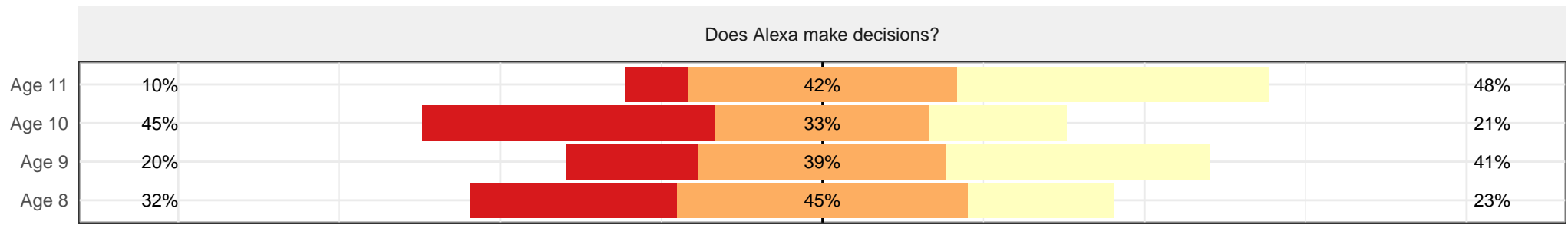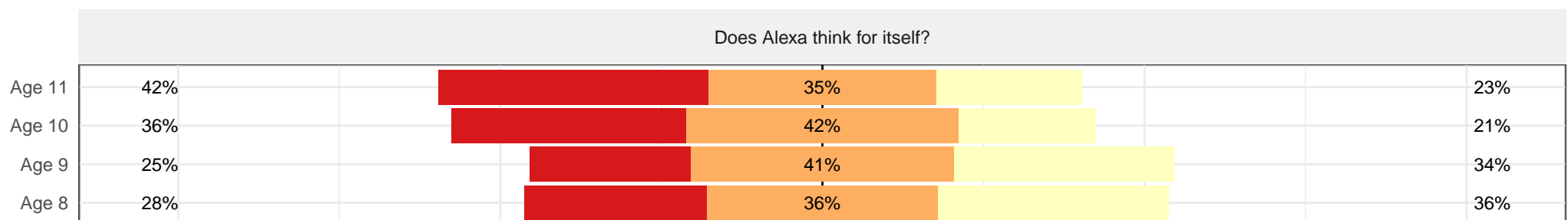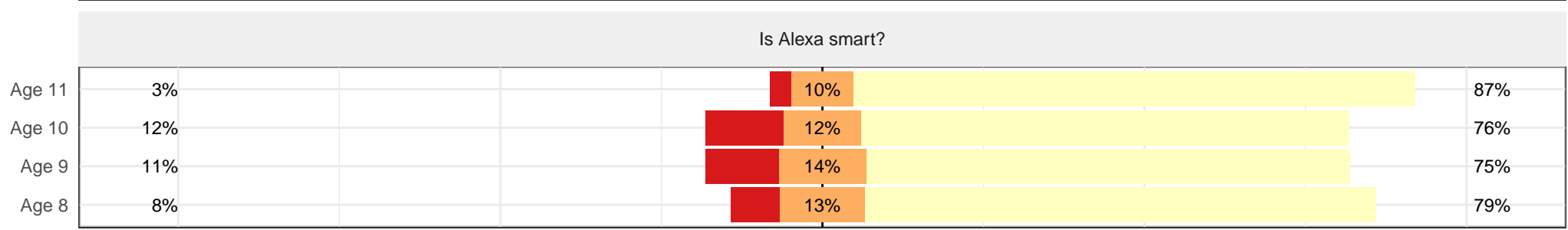

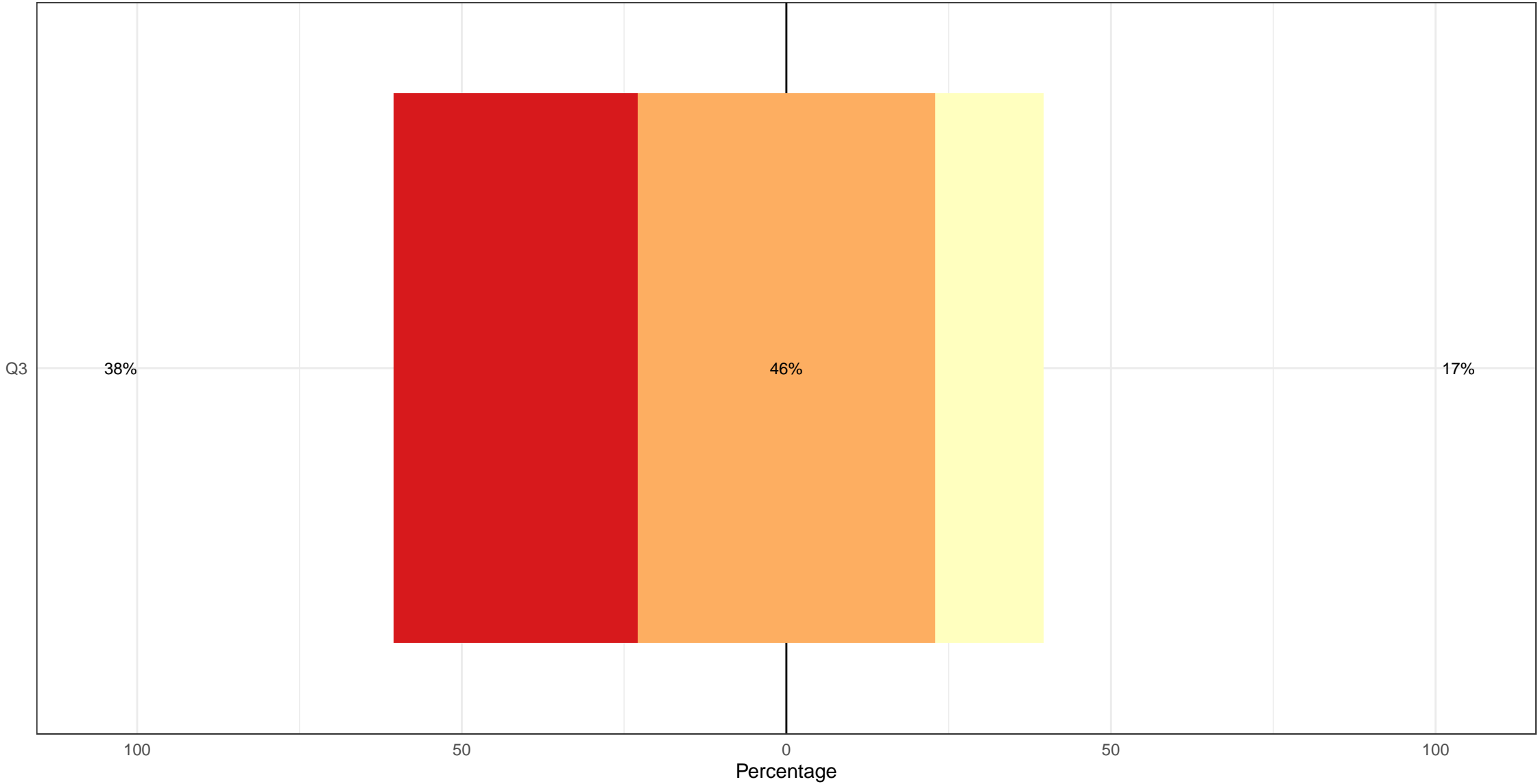

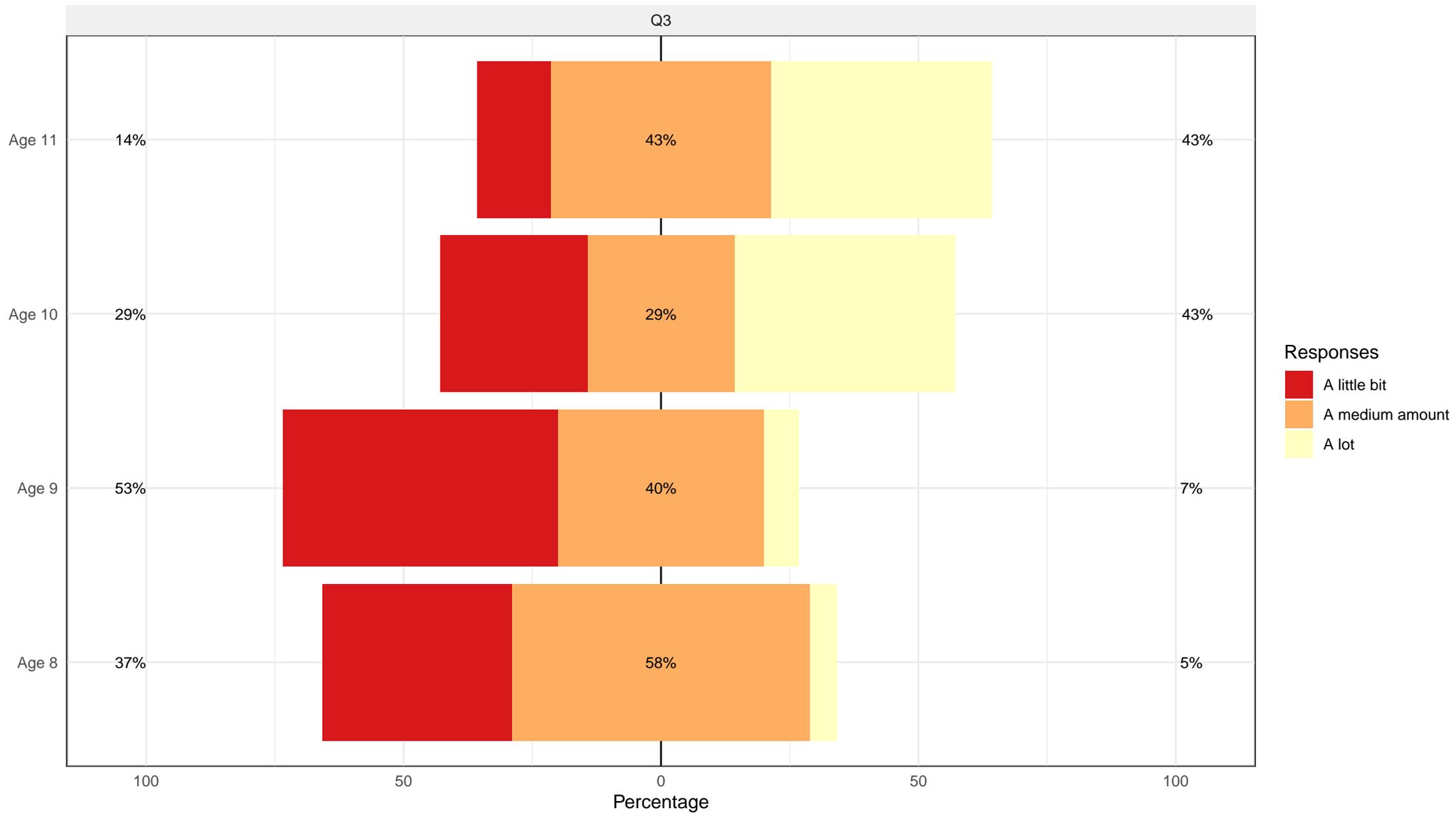

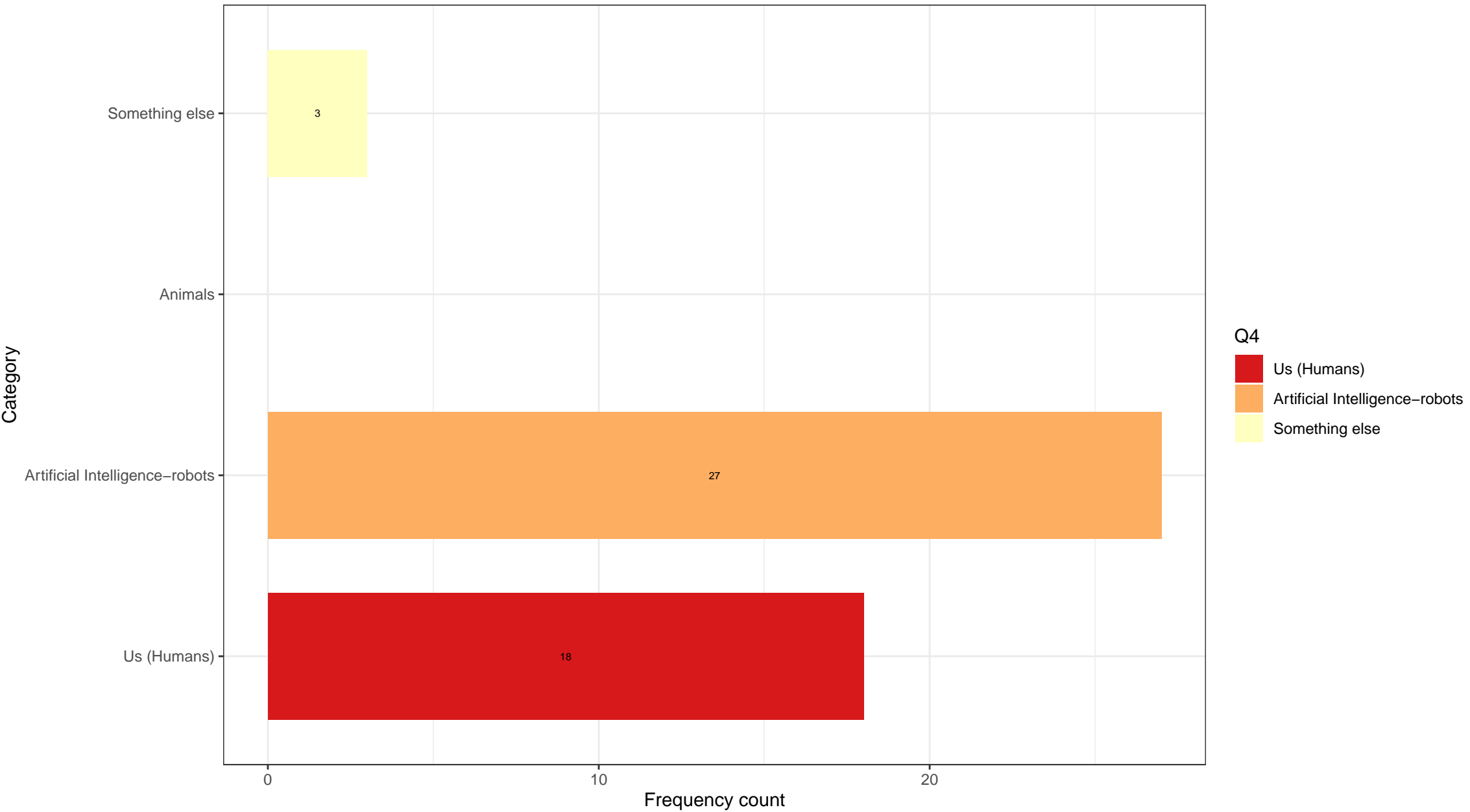

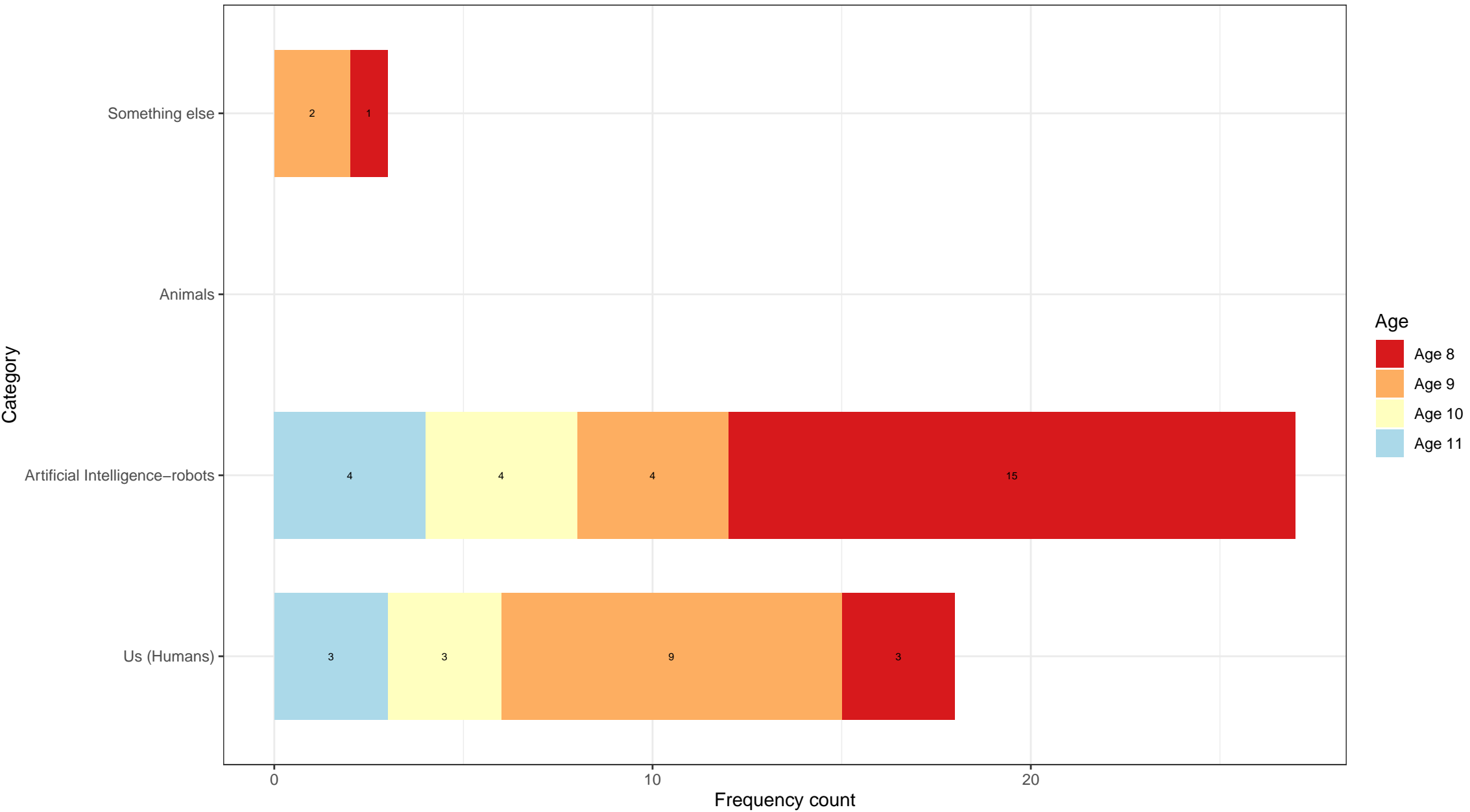

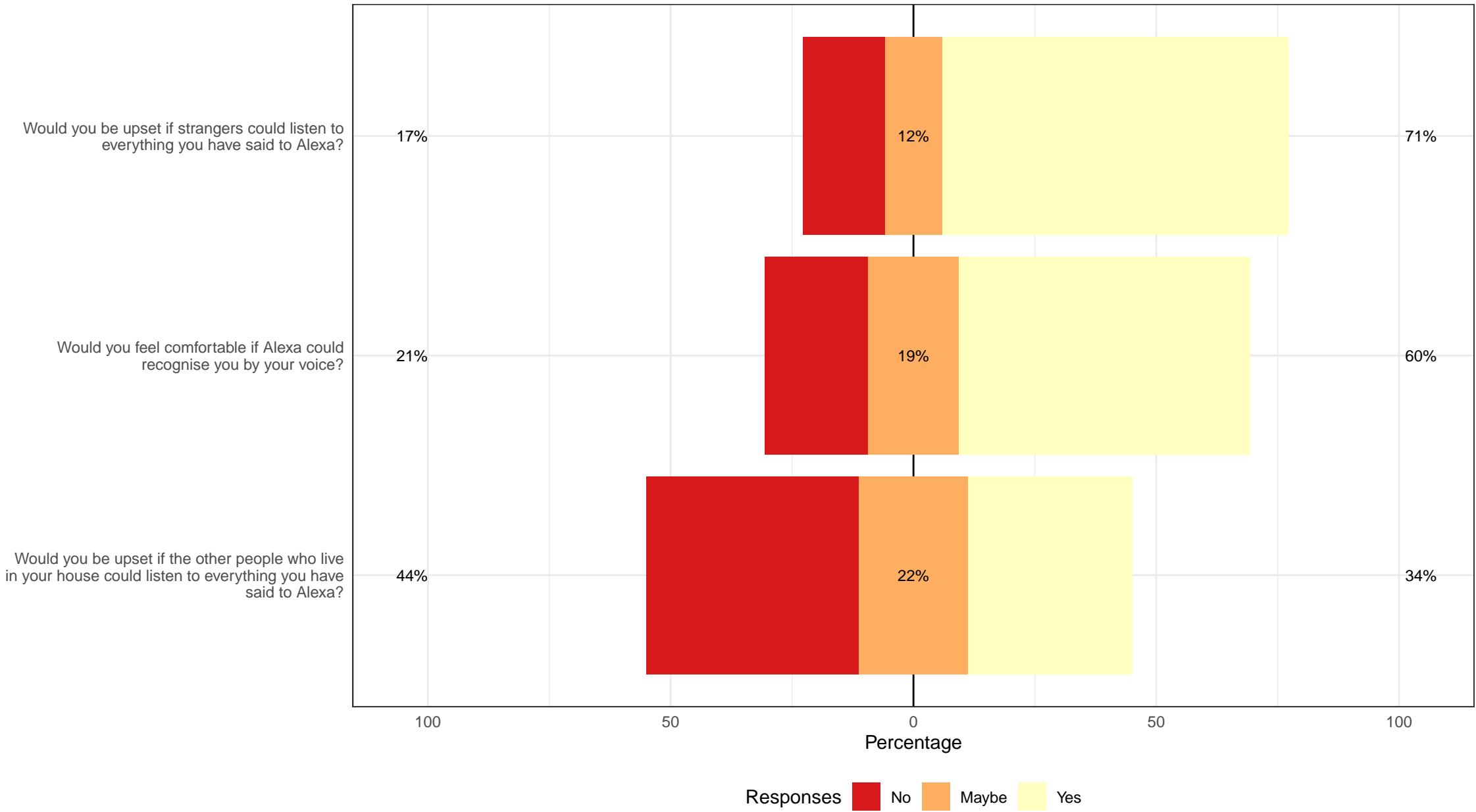

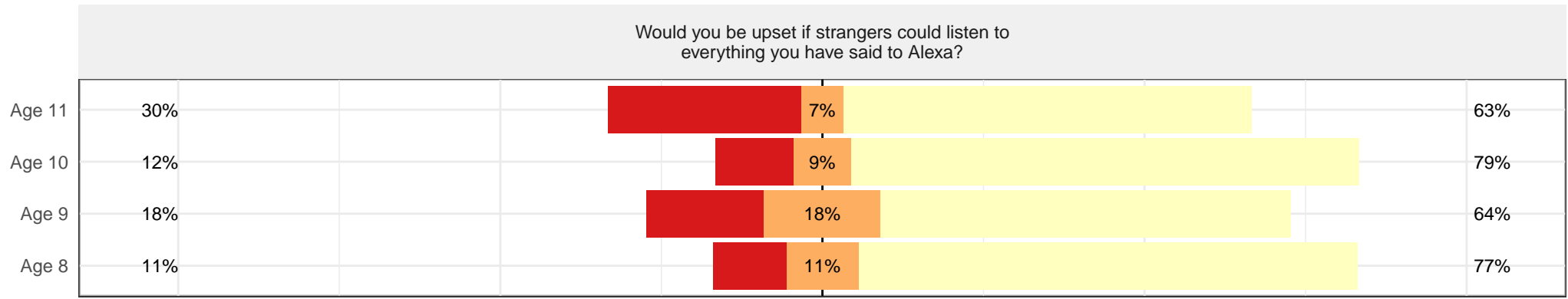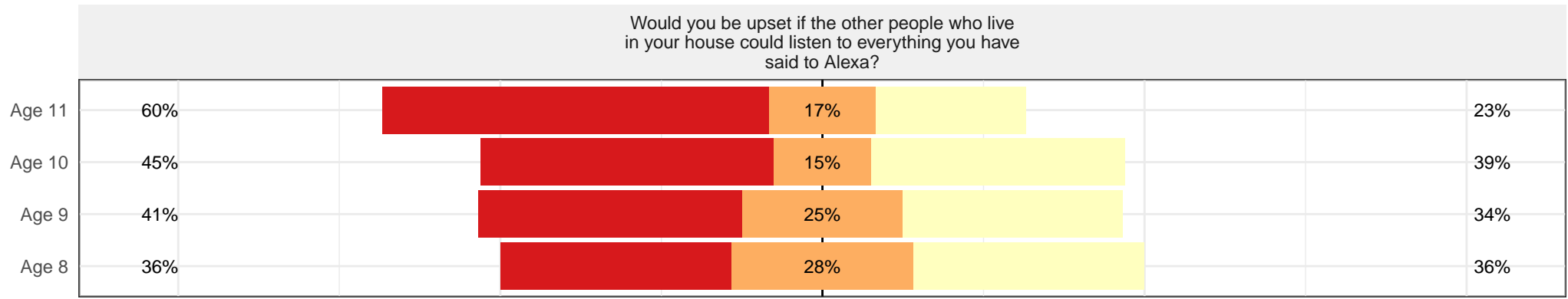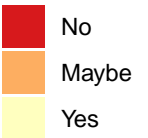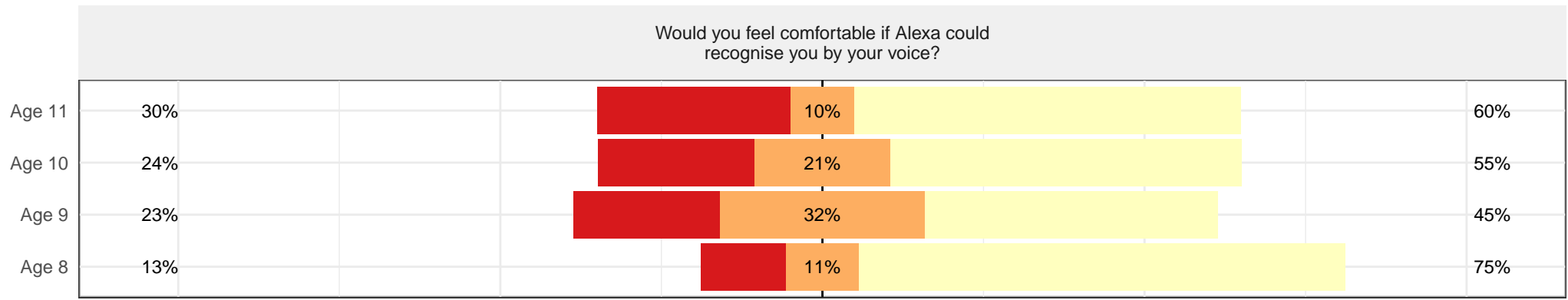

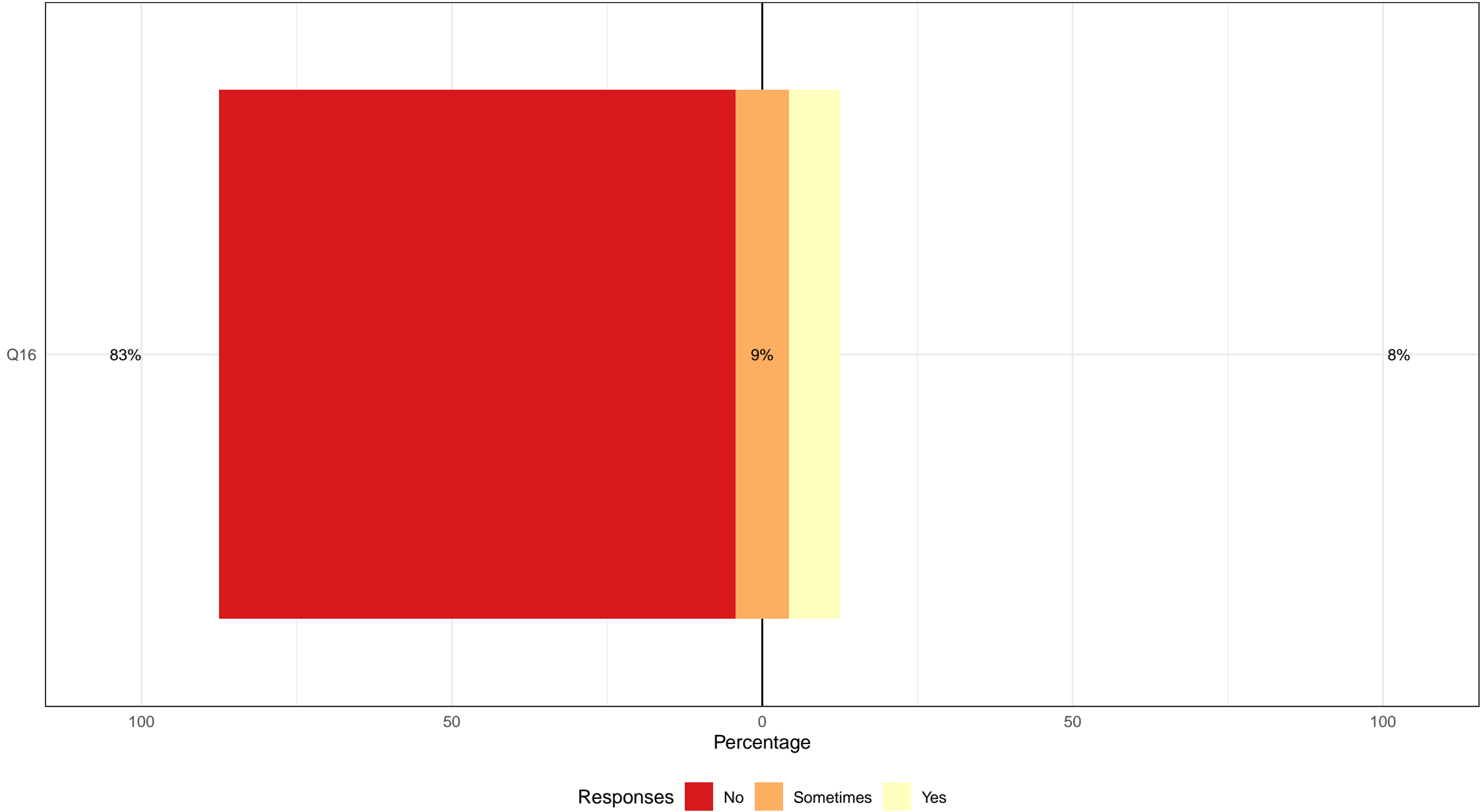

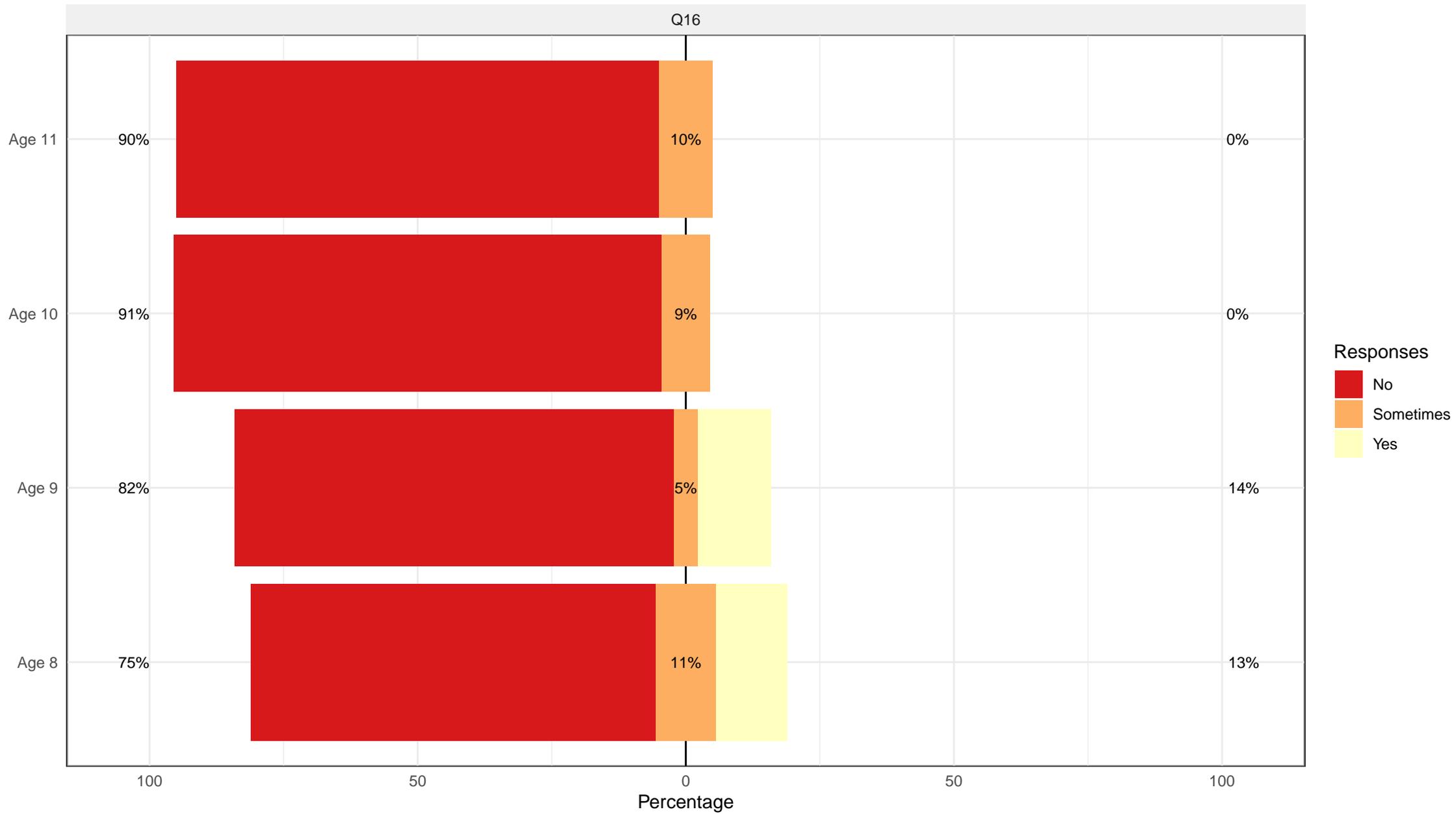

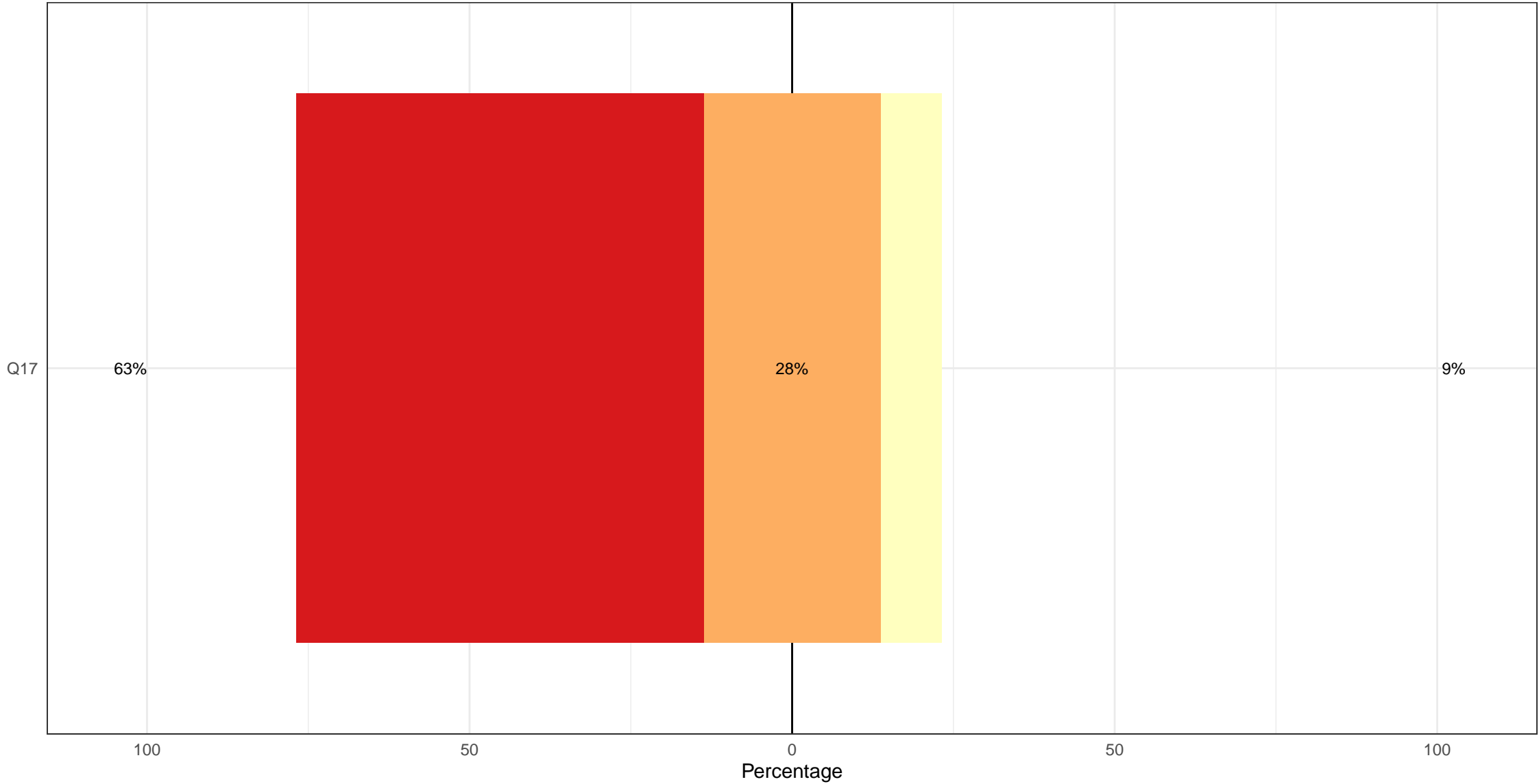

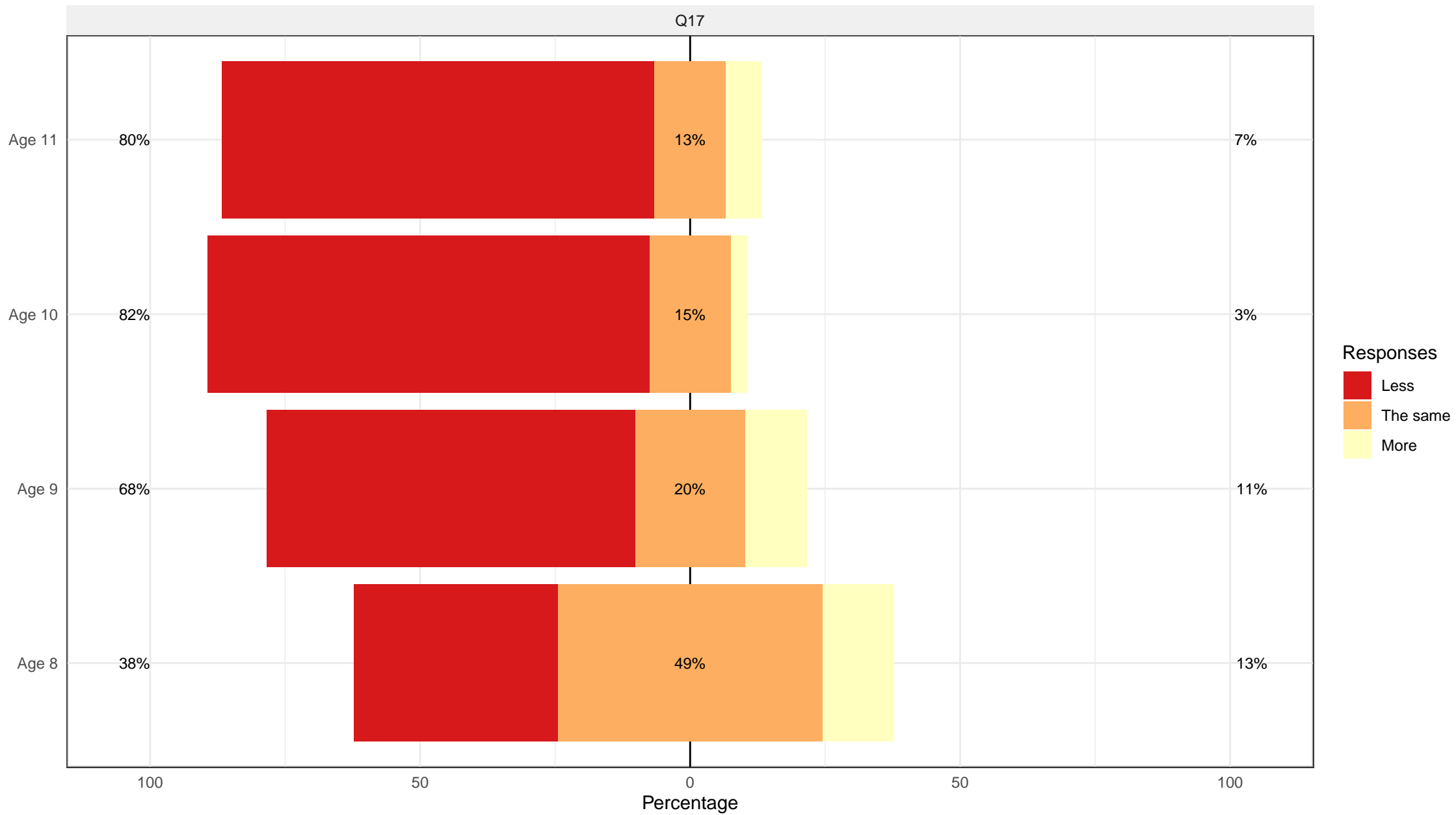

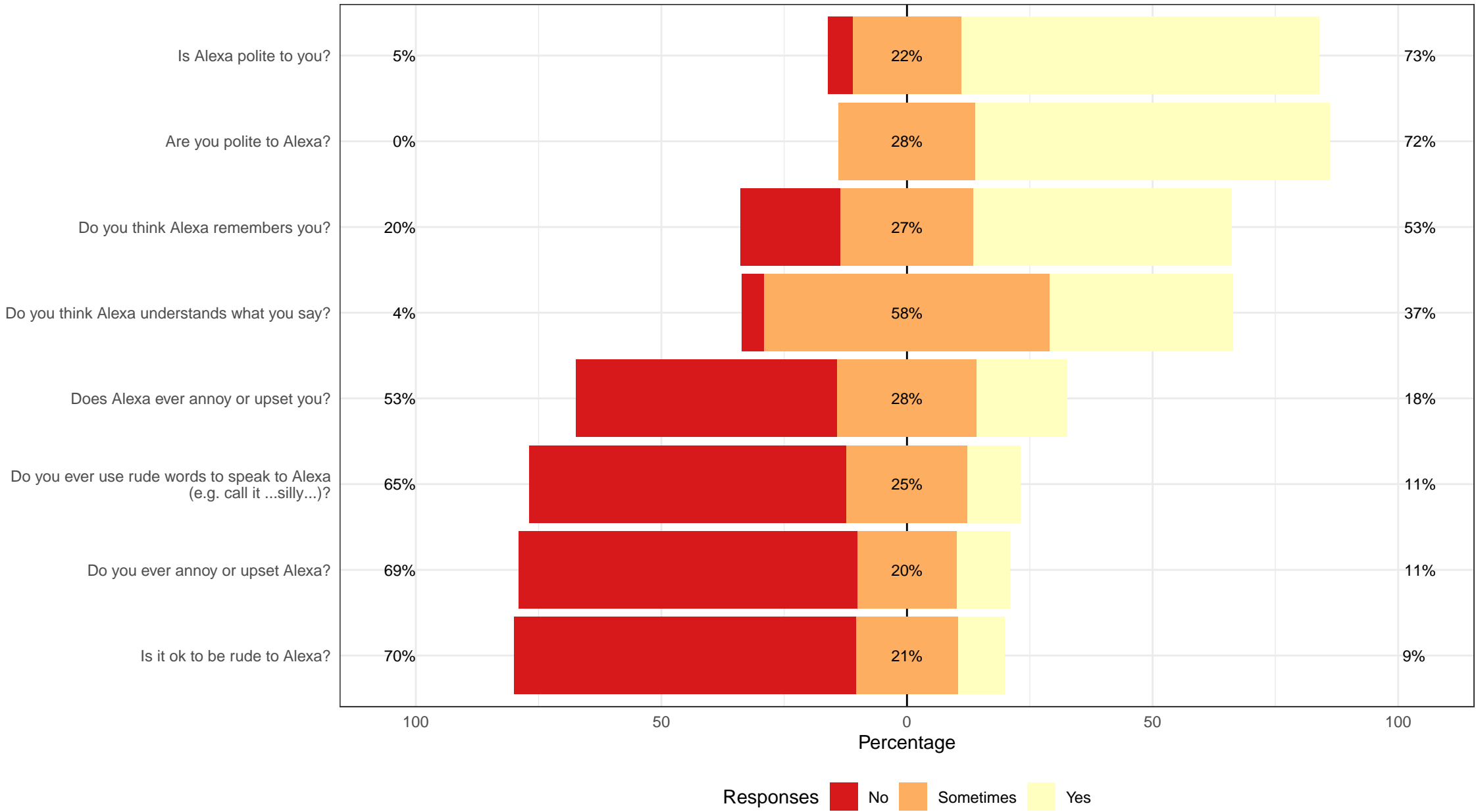

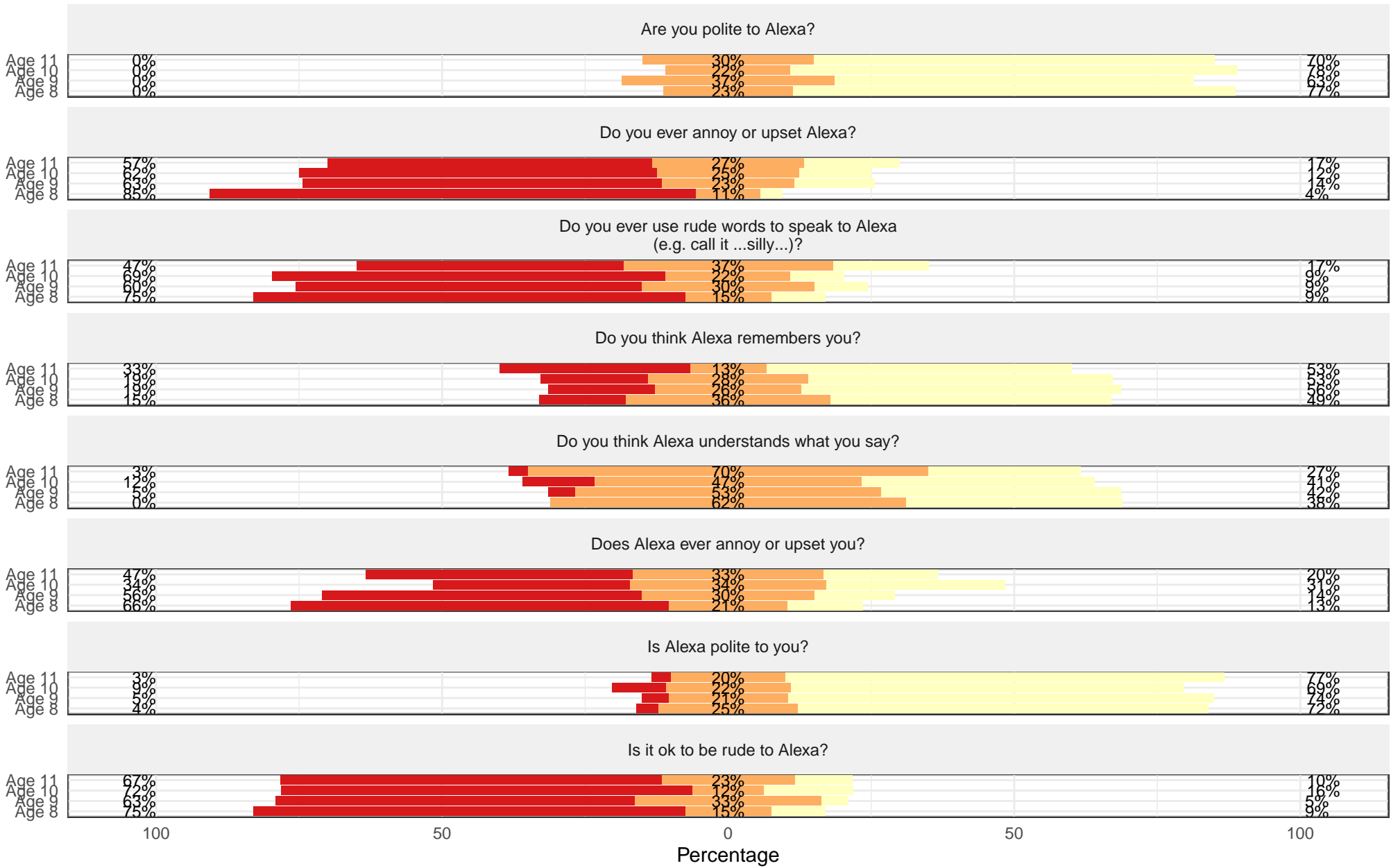

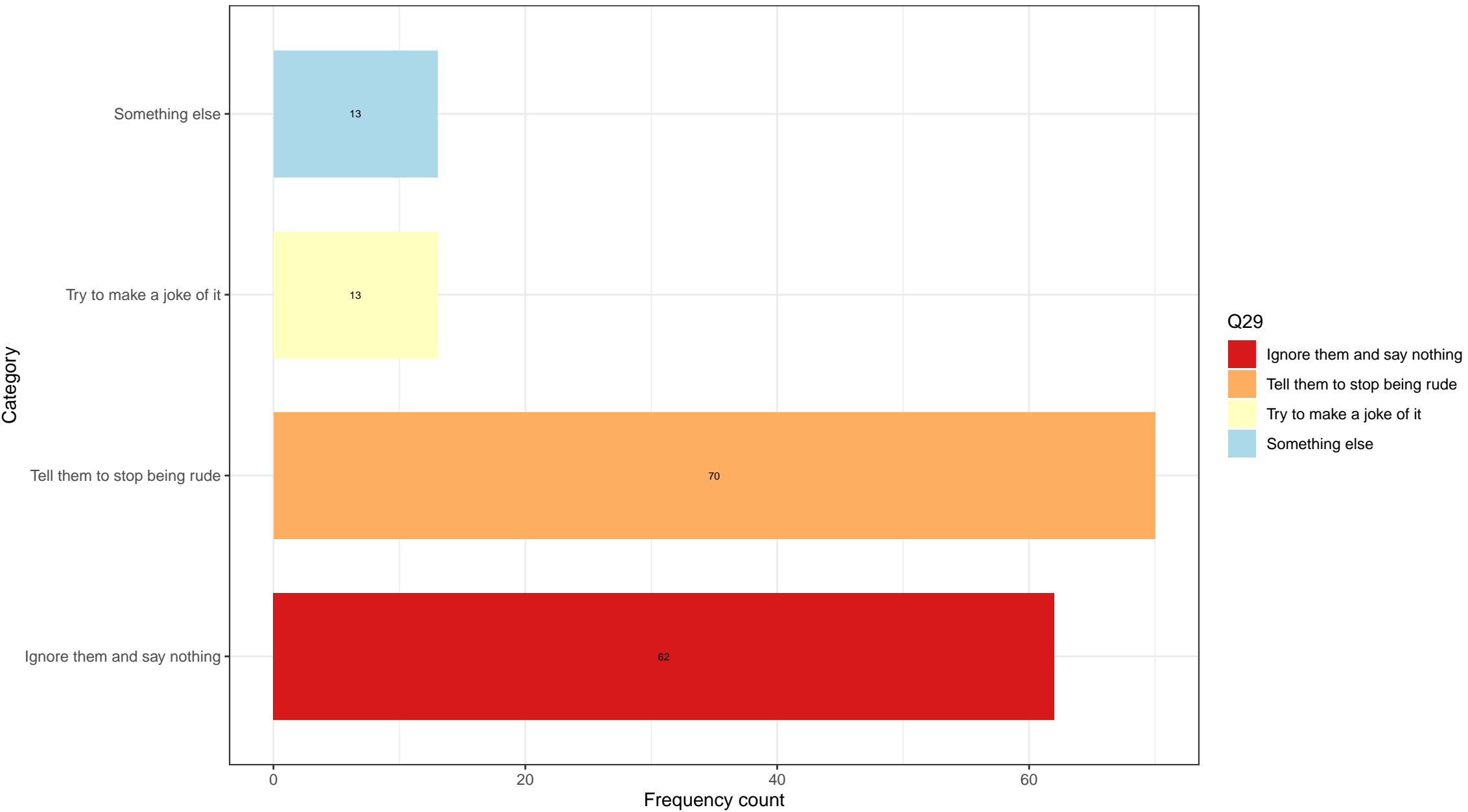

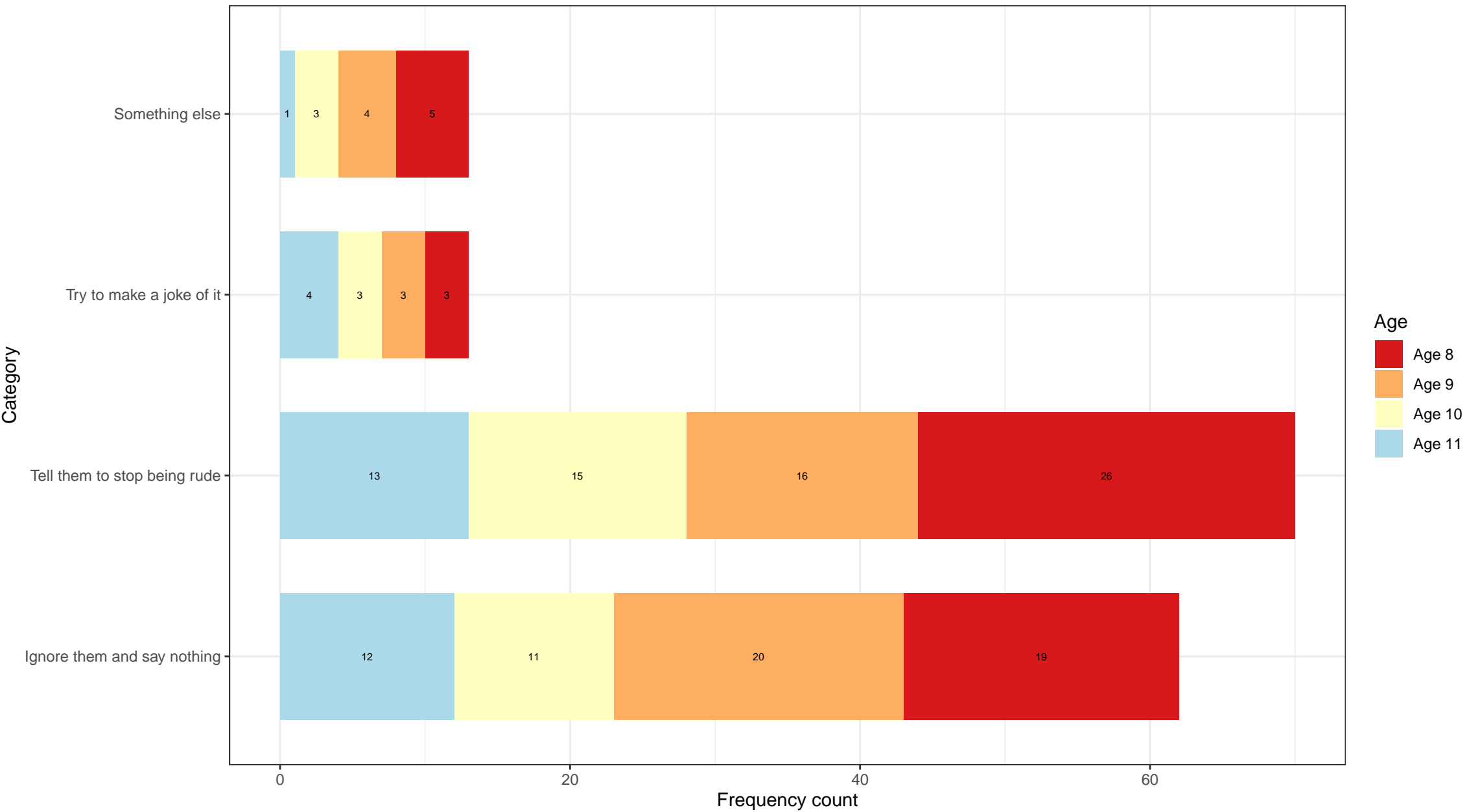